\documentclass[a4paper]{spie}  
\usepackage[]{graphicx}
\usepackage{amstext} 
\usepackage{amssymb} 
\usepackage{amsmath}
\usepackage{txfonts}
\usepackage{hyperref} 
\usepackage{url} 

\usepackage{caption}
\usepackage{subcaption}

\hypersetup{bookmarks=true,colorlinks=true,linkcolor=blue,citecolor=blue,urlcolor=blue}

\def\degr{\hbox{$^\circ$}}
\newcommand*\arcmin{\ensuremath{^\prime}}
\newcommand*\sun{\ensuremath{\odot}}
\newcommand*\farcs{\ensuremath{\overset{\prime\prime}{.}}}
\newcommand*\farcm{\ensuremath{\overset{\prime}{.}}}

\title{Enabling science with Gaia observations of naked-eye stars}
\author{J. Sahlmann\supit{a,b}, J. Mart\'in-Fleitas\supit{b,c}, A. Mora\supit{b,c}, A. Abreu\supit{b,d}, C. M. Crowley\supit{b,e}, E. Joliet\supit{b,f} 
\skiplinehalf
\supit{a}European Space Agency, STScI, 3700 San Martin Drive, Baltimore, MD 21218, USA;\\
\supit{b}European Space Agency, ESAC, P.O. Box 78, Villanueva de la Ca\~nada, 28691 Madrid, Spain;\\
\supit{c}Aurora Technology, Crown Business Centre, Heereweg 345, 2161 CA Lisse, The Netherlands;\\
\supit{d}Elecnor Deimos Space, Ronda de Poniente 19, Ed. Fiteni VI, 28760 Tres Cantos, Madrid, Spain;\\
\supit{e}HE Space Operations BV, Huygensstraat 44, 2201 DK Noordwijk, The Netherlands;\\
\supit{f}California Institute of Technology, Pasadena, CA, 91125, USA}
\authorinfo{Further author information: (Send correspondence to J.S.)\\J. S.: E-mail: johannes.sahlmann@esa.int}
\begin{document} 
\maketitle 

%%%%%%%%%%%%%%%%%%%%%%%%%%%%%%%%%%%%%%%%%%%%%%%%%%%%%%%%%%%%% 
\begin{abstract}
ESA's \emph{Gaia} space astrometry mission is performing an all-sky survey of stellar objects. At the beginning of the nominal mission in July 2014, an operation scheme was adopted that enabled \emph{Gaia} to routinely acquire observations of all stars brighter than the original limit of $G$$\sim$6, i.e.\ the naked-eye stars. Here, we describe the current status and extent of those observations and their on-ground processing. We present an overview of the data products generated for $G$$<$6 stars and the potential scientific applications. Finally, we discuss how the \emph{Gaia} survey could be enhanced by further exploiting the techniques we developed.

\end{abstract}

%%%%%%%%%%%%%%%%%%%%%%%%%%%%%%%%%%%%%%%%%%%%%%%%%%%%%%%%%%%%%
\section{INTRODUCTION}\label{sec:intro}
There are about 6000 stars that can be observed with the unaided human eye. Greek astronomer Hipparchus used these stars to define the magnitude system still in use today, in which the faintest stars had an apparent visual magnitude of 6. These `naked-eye stars` are fundamental benchmarks for stellar astrophysics, because they can be studied in most detail. The European Space Agency's (ESA) \emph{Gaia} mission (\href{http://www.cosmos.esa.int/web/gaia/}{http://www.cosmos.esa.int/web/gaia/}) is currently performing a revolutionary survey of about one billion stars in our Galaxy. \emph{Gaia} is sensitive down to magnitude $G\sim 20$ and although the observation system uses adaptive exposure times to observe stars of different magnitudes, the available dynamic range led to an original bright limit of $G\simeq6$, thus leaving out the naked-eye stars.

We are pursuing several strategies to overcome this bright limit and to secure astrometric, photometric, and spectroscopic measurements of naked-eye stars with \emph{Gaia}. In our 2014 contribution (Ref.\ \citenum{Martin-Fleitas:2014aa}), we showed how we can enable these observations and presented preliminary results obtained during the spacecraft commissioning. Here, we present the continuation of these efforts with a focus on the data that is being collected and how they can be used for scientific exploitation. We update on the results obtained during the ongoing nominal mission phase and discuss subsequent developments that can lead to an optimised scientific output. We present the motivation for this work in Sect.\ \ref{sec:science} and describe briefly the \emph{Gaia} observation system in Sect.\ \ref{sec:gaia}. Then we discuss the bright limit for nominal observations (Sect.\ \ref{sec:brightLim}), our implementation of a special observing mode for extremely bright stars (Sect.\ \ref{sec:SIF}), and the prospects for a more powerful mode (Sect.\ \ref{sec:vo}).

\section{SCIENCE OPPORTUNITIES FOR GAIA OBSERVATIONS OF $G<6$ STARS}\label{sec:science}
We present a non-exhaustive list of science cases for \emph{Gaia} astrometric observations of stars brighter than the nominal magnitude limit, here named \emph{very bright stars} or naked-eye stars. The aim of this exercise is threefold: (1) Make the community aware of the data products that potentially will be made available as part of a \emph{Gaia} data release. (2) Identify areas where ancillary observations collected by independent observers and observatories contemporaneously with \emph{Gaia} can significantly enhance the scientific output. (3) Provide scientific motivation to pursue and enhance the acquisition and analysis of very bright star data. 

\begin{figure}[h!]
\begin{center}
\includegraphics[height = 0.39\linewidth]{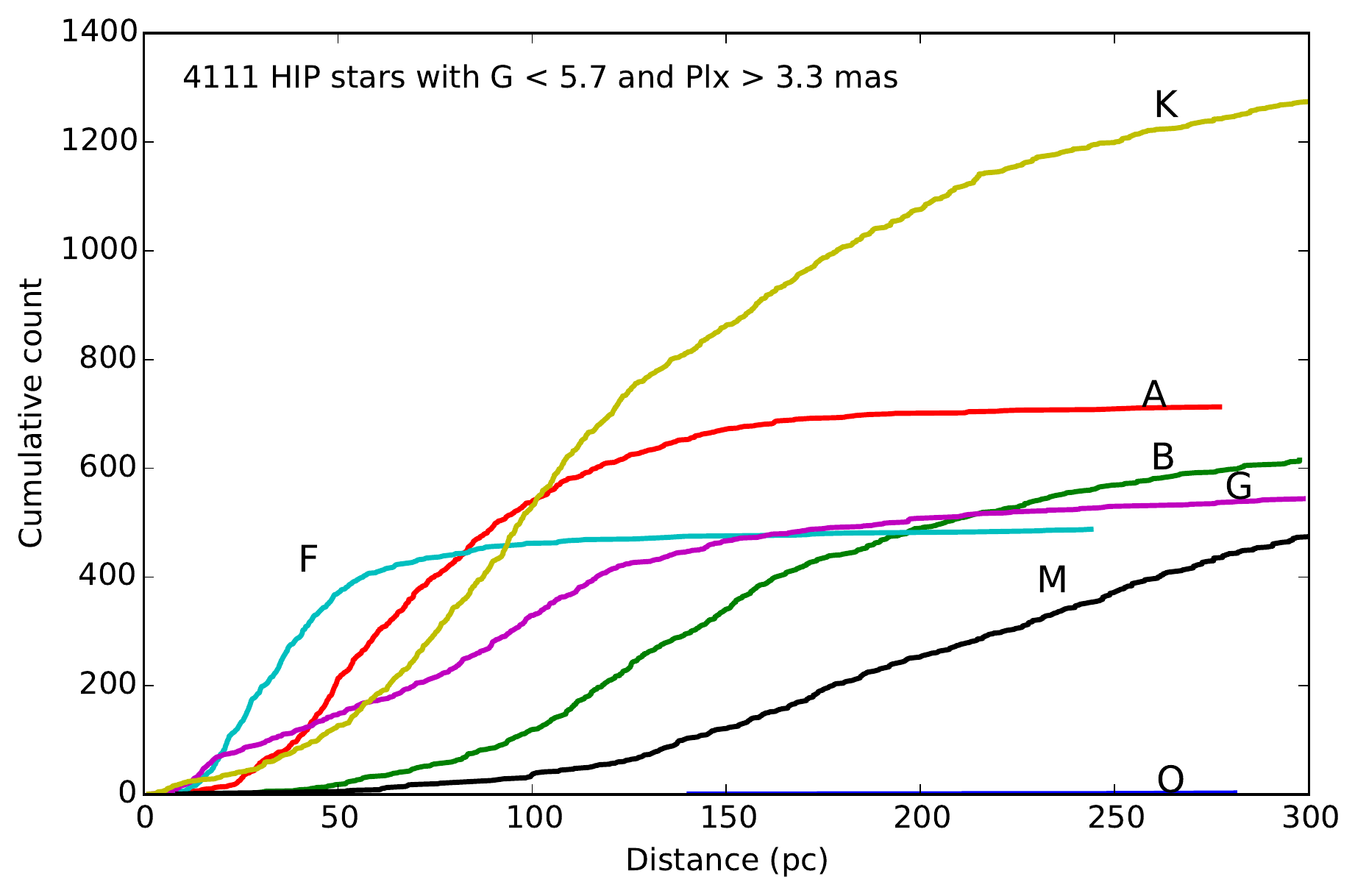} 
\includegraphics[height = 0.38\linewidth]{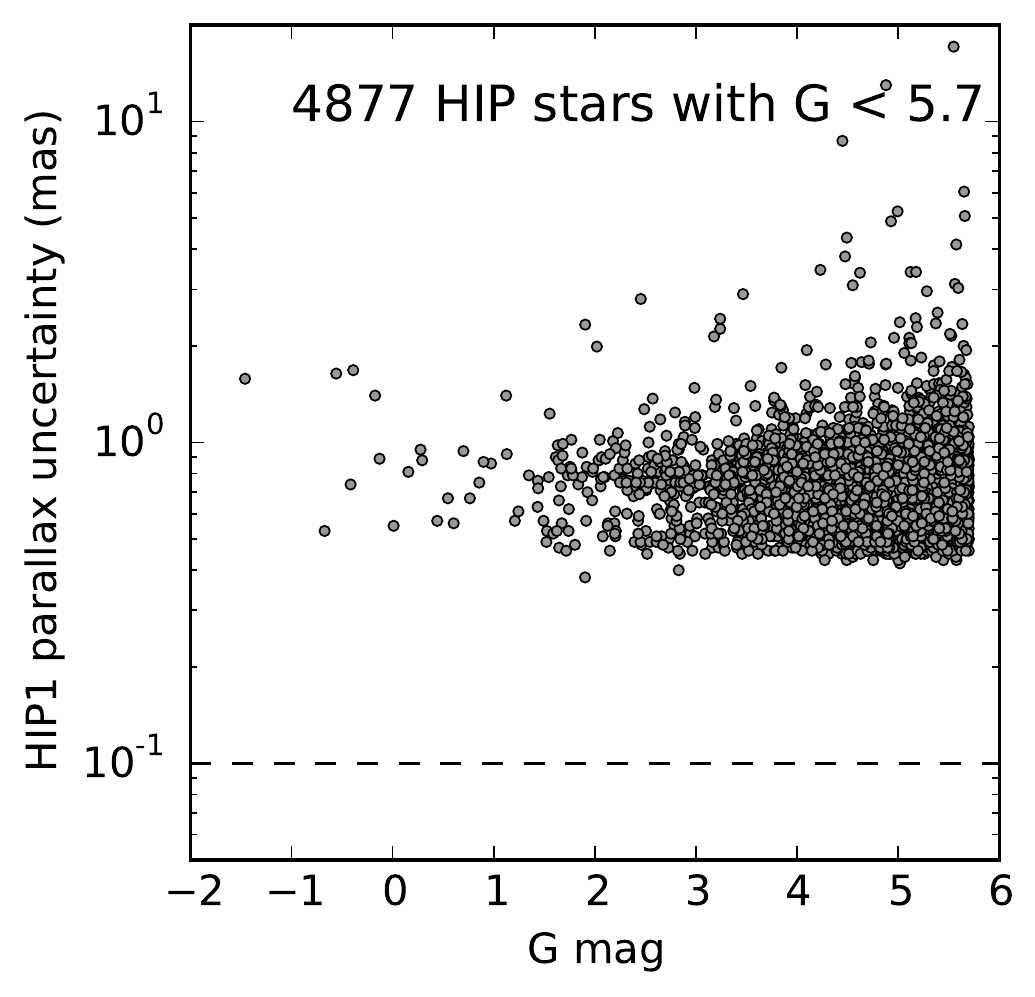}
\caption{\emph{Left}: Cumulative histograms for very bright stars closer than 300 pc for different \emph{Hipparcos}-assigned spectral types, which are indicated for each curve by a character. \emph{Right}: \emph{Hipparcos} parallax uncertainties as a function of $G$ magnitude for very bright stars. The dashed line corresponds to 0.1 mas. }
\label{fig:2}\end{center}\end{figure}

We base this section mainly on the original \emph{Hipparcos} catalogue \cite{ESA:1997vn}. A base catalogue was generated by selecting all \emph{Hipparcos} stars with $V$$\,<\,$7 and the $V$--$I$ colours of these bright stars are used to estimate their broadband \emph{Gaia} $G$ magnitude following Ref.\ \citenum{Jordi:2010kx}. We assumed that the {very bright} cut-off magnitude is $G=5.7$, because \emph{Gaia}'s onboard detection algorithm was designed to be efficient for stars up to $G=5.7$ to ensure completeness at $G=6$. For the \emph{Gaia} performance on very bright stars we usually considered two scenarios with single-measurement precisions of 100 and 50~$\mu$as. 

\subsection{Accurate masses of known exoplanets}
Most of the known exoplanets around very bright stars were discovered with radial velocity (RV) measurements and are not transiting their host star (the brightest known transiting planet host star HD 219134\cite{Motalebi:2015aa} has $V\sim5.6$). The deduced planet masses are therefore uncertain due to the unknown orbital inclination, named the $\sin i$ ambiguity. Astrometric measurements of the stellar barycentric orbit can determine the orbit's inclination accurately and remove the mass ambiguity \cite{Zucker:2001ve, Benedict:2010ph, Sahlmann:2011lr,Sahlmann:2011fk}, thereby leading to a better determination of the planet mass distribution. Because 5 of the 7 parameters describing the orbit are known from RV, the astrometric orbit detection is eased\cite{Sahlmann:2012fk2}. For a given RV planet system, we require $\mathrm{S/N}=a_1\cdot (\sigma_{\Lambda}/\sqrt{N_\mathrm{Gaia}})^{-1} > 8$ for a robust detection\cite{Sahlmann:2011fk}, where $a_1$ is the semi-major axis of the detected orbit, $\sigma_{\Lambda}$ is the single-measurement precision, and $N_\mathrm{Gaia}$ is the number of independent astrometric measurements, assumed to be 70. Because $\sqrt{70}\approx 8$, the S/N requirement translates to $a_1 \gtrsim \sigma_\Lambda$ in this case.
We queried the \href{http://exoplanets.org}{exoplanets.org}\cite{Wright:2011lr} database on February 18, 2016, and retrieved systems with entries for planet mass, star mass, parallax, and orbital period. For each system, we computed the semimajor axis $a_1 = a_{1,min}$, which usually is a lower limit because we set $M_2 \sin i = M_2$ when $i$ is unknown, which is mostly the case. We imposed a period limit of 3000 days to ensure that \emph{Gaia} measurements during the nominal 5-year mission lifetime cover a large ($\gtrsim 66$\%) portion of the orbit. The results are summarised in Table~\ref{tab:2} and the individual systems are listed in Table \ref{tab:1a}. Provided a precision of 100 $\mu$as can be reached for stars with magnitudes $G=5.7-4.5$, 13 known exoplanet orbits will be fully characterised, i.e.\ the planet masses will be determined accurately. If 50 $\mu$as can be reached up to $G=2$, the number of such systems increases to 38. 
\setcounter{table}{0}
\begin{table}[h!]
\caption{Number of characterisable RV planets around very bright stars as a function of precision and star magnitude.\vspace{2mm}}
\label{tab:2} 
\centering  
\begin{tabular}{c | c c c} 
\hline\hline 
Single-measurement & \multicolumn{3}{c}{$G$ magnitude}\\
precision ($\mu$as) & $5.7-4.5$ & $4.5-2.0$ & $<$2.0  \\
\hline
 50 -- 100 & 7 & 9 & 1  \\
100 & 13 & 7 & 3  \\
 \hline
\end{tabular} 
\end{table}

\subsection{Discovering new exoplanets around very bright stars}
Extending the \emph{Gaia} magnitude range towards brighter stars gives access to thousands of nearby stars that can be searched for astrometric signatures of yet undetected extrasolar planets. A key advantage of such a uniform astrometric survey is that it will be able to detect giant planets around a variety of stars regardless of their evolutionary state, including fast rotators and young and active stars, which are difficult to target with other techniques.
About 4870 stars in the original \emph{Hipparcos} catalogue (with available $V-I$ colour) are brighter than $G=5.7$. Of these, 66 stars are ultra-bright with $G<2.0$. Because the astrometric detection capability decreases reciprocally with distance, we studied stars with parallax $>3.3$ mas, i.e.\ closer than 300 pc from the Sun, which leaves about 4100 stars. Figure~\ref{fig:2} shows these stars roughly grouped by spectral type and Table~\ref{tab:5} shows the number of stars per spectral type, distance, and magnitude bin.\\ 
Because these stars will be searched `blindly` for the presence of extrasolar planets, the minimum required S/N for detection is $\sim$20\cite{Sahlmann:2015aa}, i.e.\ much higher than when the RV orbit is known. Assuming a stellar mass of 1 Solar mass ($M_{\sun}$) and a precision of 50 $\mu$as for 70 measurements over 5 years, a one Jupiter mass ($M_J$) planet in a 2000 day orbit can then be detected out to a distance of 20 pc and a 10 $M_J$ planet can be detected around a star up to 200 pc away. 

\begin{table}[h!] 
\small
\caption{Number of very bright stars within 300 pc as a function of spectral type, luminosity class, distance, and magnitude. The total number is 4111. The row labelled `Any` lists the sum of the respective column. For spectral type and luminosity class of multiple stars, we used the properties of the primary component. Stars for which a luminosity class is not available in the original catalogue\cite{ESA:1997vn} are listed in the `N/A` column. \vspace{2mm}}
\label{tab:5} 
\centering  
\begin{tabular}{c | c c c c c c | ccc | ccc } 
\hline\hline 
Spectral type & \multicolumn{6}{c}{Luminosity class} & \multicolumn{3}{c}{Distance (pc)} & \multicolumn{3}{c}{$G$ magnitude range}\\
 & I & II & III & IV & V & N/A & $0-50$ & $50-100$ & $100-300$ &$5.7-4.5$ & $4.5-2.0$ & $<$2.0\\
\hline
O & 1 & 1 & 0 & 0 & 1 & 0 &0 &0 &3 &0 &2 &1  \\
B & 6 & 12 & 98 & 132 & 315 & 52 &18 &101 &496 &409 &189 &17  \\
A & 2 & 1 & 56 & 97 & 399 & 158 &214 &324 &175 &541 &162 &10  \\
F & 2 & 19 & 70 & 103 & 260 & 34 &370 &92 &26 &381 &101 &6  \\
G & 15 & 50 & 314 & 41 & 70 & 54 &147 &182 &215 &402 &141 &1  \\
K & 10 & 59 & 1040 & 30 & 25 & 111 &127 &403 &744 &966 &291 &17  \\
M & 4 & 16 & 375 & 1 & 3 & 74 &5 &31 &438 &303 &161 &10  \\
Any & 40 & 158 & 1953 & 404 & 1073 & 483 &881 &1133 &2097 &3002 &1047 &62  \\
\hline 
\end{tabular} 
\end{table}

We used this criterion to qualitatively assess the \emph{Gaia} discovery potential as a function of spectral type: There are only 3 very bright O-type stars, which is negligible for this discussion. For very bright M stars, we identify a small potential because most are far-away M-type giants and the frequency of close giant planets is known to be low\cite{Bonfils:2013aa, Montet:2014ab}. B stars are most numerous beyond 100 pc, which lowers the probability of detecting planetary companions. However, the population of close planets around these stars is uncertain\cite{Janson:2011fk}. A large population of very massive planets ($10-40\,M_J$) in close orbits ($\lesssim5$AU) could be detectable by \emph{Gaia}. For G and K dwarf stars, the close planet population statistics are well studied with RV\cite{Mayor:2011aa}. However, \emph{Gaia} can refine our knowledge of giant planets in intermediate-period orbits and also discover planets around very bright G/K giant stars.
The probably greatest potential is offered by the A-F type stars, of which hundreds with $G<5.7$ are located within 50 pc. The planet population around stars with spectral types A and earlier than mid-F is poorly known, mainly due to sensitivity limitations of RV caused by stellar activity and rotation\cite{Lagrange:2009lr}. The Frequency of Jovian planets around (evolved) A-F type stars within 3 AU is estimated at $26\pm9$ \%\cite{Bowler:2010fk}. We can therefore expect to detect tens of new extrasolar planets around nearby stars in a parameter space that is difficult to access with techniques other than \emph{Gaia} astrometry.

\subsection{Parallaxes and proper motions of very bright stars}
Assuming a single-measurement precision of 100 $\mu$as, we expect an end-of-mission parallax accuracy of at least the same order, thus the decrease in parallax uncertainty compared to \emph{Hipparcos} for very bright stars will be about one order of magnitude, see Figure~\ref{fig:2}. \emph{Gaia} is the only instrument able to deliver such measurements in the foreseeable future and we expect those to be relevant for a large number of science cases. The only scheduled space mission for astrometry of bright stars is nano-Jasmine\cite{Yamada:2013aa}, which aims at $\sim$3 mas accuracy.\\ 
Exoplanet research is a core theme for the next major space observatories JWST and WFIRST, and for proposed astrometric missions like NEAT\cite{Malbet:2012aa}/\emph{Theia}. 
The astrometric characterisation of the very bright stars with \emph{Gaia} will yield precision parallaxes and proper motions and, more importantly, will reveal the presence of giant planets, which is invaluable for defining and optimising the target lists for these missions. Stars that exhibit an astrometric acceleration term compatible with a wide planetary companion are also ideal targets for ground/space-based exoplanet imagers like GPI and  SPHERE or JWST-NIRISS\cite{artigau2014}. The proximity and brightness of these stars both play in favour of subsequent direct imaging confirmation and characterisation of the planets.

\subsection{Binary stars}
Binary stars will be present in the sample of very bright stars and become apparent as astrometric binaries in the \emph{Gaia} measurements. To estimate the expected amount of binaries, we inspected the astrometric solution types of very bright stars given in the new \emph{Hipparcos} catalogue \cite{:2007kx}. As shown in Table~\ref{tab:4}, we can expect non-standard astrometric behaviour for at least 25 \% of very bright stars. Because of the increased sensitivity of \emph{Gaia} compared to \emph{Hipparcos}, we expect to extend the astrometric binary study of very bright stars towards smaller secondary-to-primary mass ratios. The use of external catalogues, e.g.\ of spectroscopic binary stars (SB9\cite{Pourbaix:2004yq}), is not further explored here, but we expect that our knowledge of known binaries will be refined with \emph{Gaia} astrometry and that several new binary systems will be discovered.
\begin{table}[h!]
\caption{Very bright stars and their HIP2 solution types.\vspace{2mm}}
\label{tab:4} 
\small
\centering  
\begin{tabular}{r r r r } 
\hline\hline 
Solution & Number of stars & \% & $S_n$\\
\hline
Standard & 4083 & 74.9 & 5 \\
Accelerated& 206 & 3.8 & 7,9 \\
Stochastic & 350 & 6.4 & 1 \\
Binary & 724 & 13.3 & 15,55,75,95 \\
Acc. binary & 82 & 1.5 & 17,57 \\

\hline % 
\end{tabular} 
\end{table}

In addition, there are many very bright stars of particular interest and \emph{Gaia} data may help us to refine our knowledge of their properties. Amongst the naked-eye stars, there are for instance several stars surrounded by disks and/or planets (e.g.\ Fomalhaut, $\beta$ Pic), the brightest members of the Pleiades cluster, stars with directly measured apparent diameters (e.g.\ Betelgeuse), the brightest Cepheids, and the brightest star of the Orion Trapezium cluster $\Theta^1$ Ori C. 

\section{GAIA OBSERVATION SYSTEM}\label{sec:gaia}
\begin{figure}[h!]
\begin{center}
\includegraphics[width = 0.8\linewidth, trim=0 0cm 0 7mm, clip]{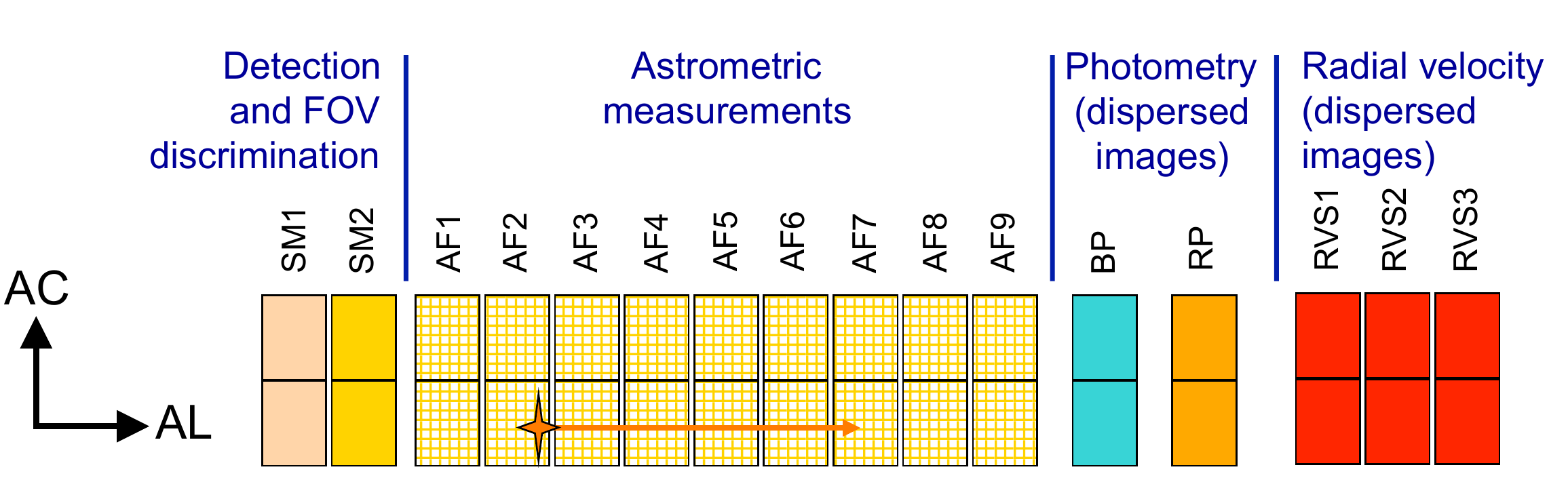} \vspace{1mm}
\caption{Schematic of two out of seven CCD rows in the \emph{Gaia} focal plane, adapted from a figure by L. Lindegren. As \emph{Gaia} spins, the image of a stellar object moves from left to right and first falls on one of the SkyMapper CCDs (SM1 or SM2, depending on the field-of-view (FOV) or, equivalently, telescope the object is observed with). It then sweeps across the astrometric field (AF1--AF9), the red and blue photometer (BP and RP), and finally the spectrograph (RVS1--RVS3) CCDs. AC and AL indicate along-scan and across-scan direction, respectively.}
\label{fig:fpa}\end{center}\end{figure}

The \emph{Gaia} spacecraft\cite{de-Bruijne:2012kx} is continuously scanning the sky with a spin period of 6 hours. It observes with two telescopes that capture two field of views separated by the basic angle of 106.5\degr, which are imaged onto a common focal plane. The focal plane array consists of 106 CCDs arranged in 7 rows and 17 functional strips (Fig. \ref{fig:fpa}). Every CCD is operated in time-delayed integration mode, where the charge is transferred as the stellar images move across the light-sensitive area. Every CCD row is controlled by a computer, the Video Processing Unit, which is in charge of the detection, confirmation, and observation of stellar objects\cite{Martin-Fleitas:2014aa}.

A stellar object crossing the focal plane is first recorded by either SkyMapper1 (SM1) or SkyMapper2 (SM2), depending on whether it is observed with telescope 1 or telescope 2, respectively. All pixels of the SkyMapper CCDs are read and the resulting images are processed in real time to determine whether the image corresponds to a star-like object\cite{de-Bruijne:2015ab}. If the outcome is positive and the object is confirmed in the first astrometric field CCD (AF1), an observation window is assigned to the object. This observation window is then propagated across all subsequent CCDs (astrometric field, photometers, and spectrograph) and only pixels falling within this window are stored on-board for download to the ground. This strategy is required to reduce the telemetry to a manageable amount. Once downlinked, the \emph{Gaia} data is fed into the processing pipeline and distributed to the data processing centres that are coordinated by the \emph{Gaia} Data Processing and Analysis Consortium (DPAC\cite{Mignard:2008aa}\ , \href{http://www.cosmos.esa.int/web/gaia/dpac}{http://www.cosmos.esa.int/web/gaia/dpac}).

\section{GAIA BRIGHT LIMIT IN OPERATION}\label{sec:brightLim}
\begin{figure}[h!]\begin{center}
\includegraphics[width = 0.45\linewidth,trim=0cm 0cm 0 0cm, clip]{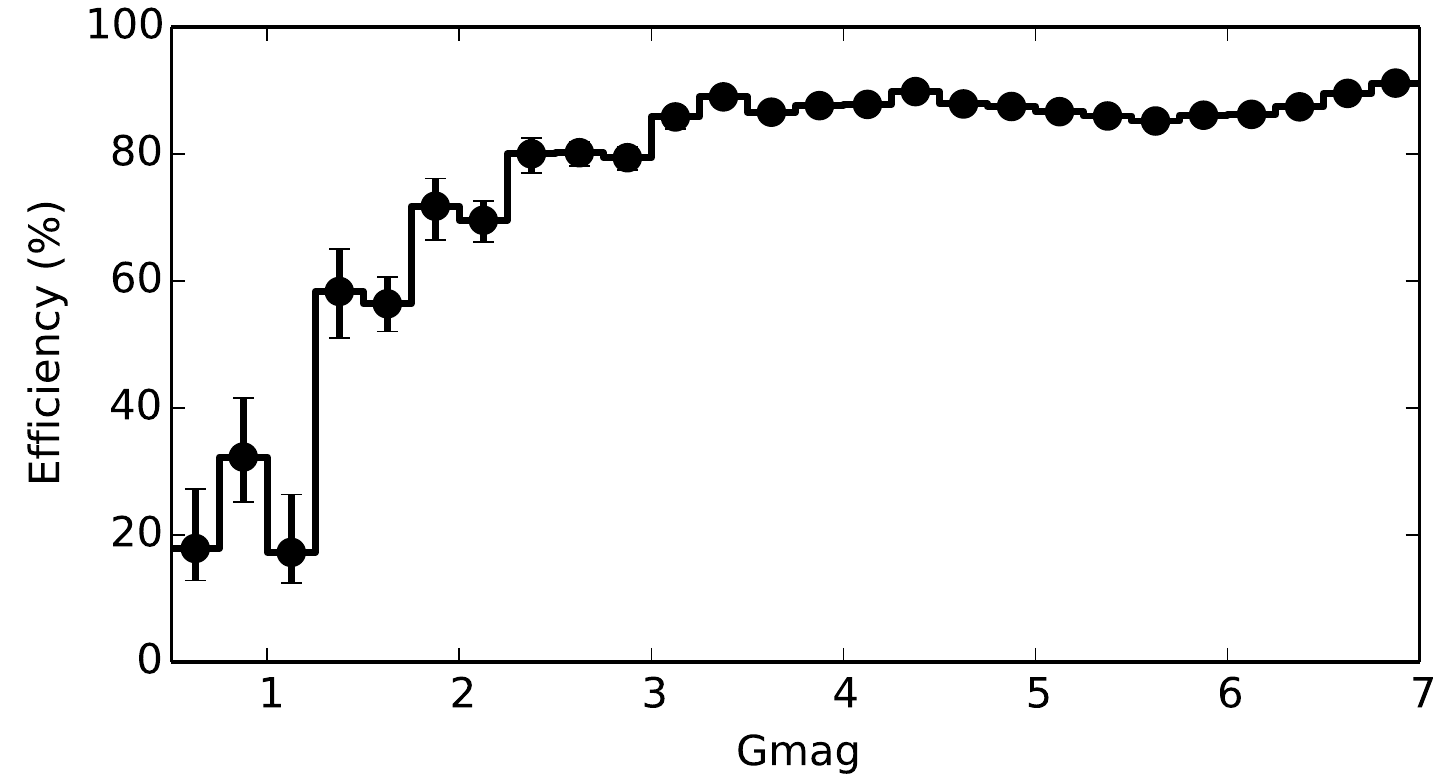} 
\includegraphics[width = 0.54\linewidth,trim=0cm 0cm 0cm 0cm, clip]{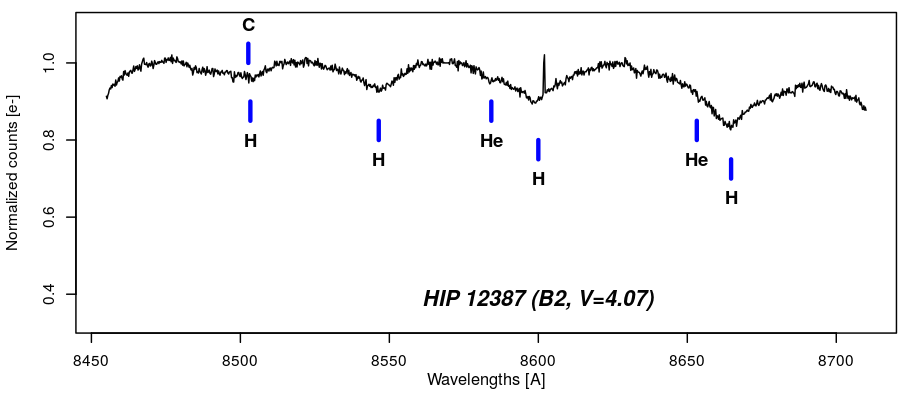} 
\caption{\emph{Left:} On-board source confirmation efficiency as a function of $G$ magnitude obtained from the analysis of nominal \emph{Gaia} observations. Black bars show the uncertainties corresponding to binomial statistics. \emph{Right:} \emph{Gaia} Radial Velocity Spectrometer (RVS) spectrum of the very bright B2 star HIP\,12387 ($V=4.1$) from nominal observations, taken from \url{http://www.cosmos.esa.int/web/gaia/iow_20141124}.}
\label{fig:effic}\end{center}\end{figure}

To investigate the actual bright limit for autonomous \emph{Gaia} observations, we compared the number of predicted focal plane crossings of bright stars with the number of corresponding \emph{Gaia} observations. As we will discuss in Sect.\ \ref{sec:vo}, we have the ability to accurately predict when and where the image of a given stellar object will cross the \emph{Gaia} focal plane, using a customised version of the publicly accessible \emph{Gaia} observing schedule tool (GOST, \href{http://gaia.esac.esa.int/gost/index.jsp}{http://gaia.esac.esa.int/gost/index.jsp}). 
We predicted the focal plane crossings of \emph{Hipparcos} stars brighter than $G<7$ using GOST for a timerange that spans \emph{Gaia} revolutions $\sim$1404 -- 2132, which corresponds to about six months starting in September 2014. We then extracted the nominal \emph{Gaia} bright star observations (in \emph{Gaia} terminology, these are class0 astroElementaries) in the same timerange from the  \emph{Gaia} main database. We crossmatched every predicted event with the nominal \emph{Gaia} observations on the basis of the observing time in the astrometric field (AF1), the field of view, and the CCD row.\\
As a result, we found 141\,800 nominal \emph{Gaia} observations that have a confirmed association to one of 161\,100 predictions. The comparison between the number of predictions in a given magnitude bin and the actual number of corresponding observations yields an estimate of the observation efficiency as a function of magnitude, which is shown in Fig.\ \ref{fig:effic}. The absolute value of the efficiency depends on the particular choices made during the analysis, e.g.\ the inclusion of data gaps with various origins and of data quality flags set during the automatic pipeline processing, which are detailed in Ref.\ \citenum{JSA-006}. In addition, we did not assess yet the scientific usability of the nominally produced data for stars brighter than $G=5.7$, although there is already very promising evidence, see Fig.\ \ref{fig:effic}. The dependence of efficiency on magnitude can be used directly to investigate the bright limit of \emph{Gaia} on-board confirmation. In broad terms, the efficiency is very high for fainter stars ($G=3-7$) and very low for extremely bright stars ($G<2$). The break point is located at $G\simeq3$, which confirms the results we obtained during \emph{Gaia} commissioning\cite{JSA-004, Martin-Fleitas:2014aa}.

\section{FORCED IMAGE DATA ACQUISITION FOR $G<3$ STARS}\label{sec:SIF}
For extremely bright stars ($G<3$) the probability of obtaining nominal \emph{Gaia} data is very low, see Fig. \ref{fig:effic}. To circumvent this, we use a special observing mode to force the data acquisition of SkyMapper images for the 230 stars brighter than $G=3$, which are listed in Tables \ref{tab:vbsstars} and \ref{tab:vbsstars2}. A detailed description of this mode is given in Ref.\ \citenum{Martin-Fleitas:2014aa}. The principle consists in using the Service Interface Function (SIF) of the Video Processing Unit to obtain all pixel data of a SkyMapper CCD during a short period of time, i.e.\ to acquire a full-frame image with the respective CCD. The resulting data are not equivalent to nominal \emph{Gaia} observations, as they are limited to SkyMapper images only, which have a larger pixel scale (due to on-chip 2$\times$2 binning) and a fixed integration time of 2.85 s (CCD gate 12).

\subsection{Prediction of focal plane crossing times}
To avoid generating unmanageable amounts of telemetry, this observing mode requires to limit the duration of an individual image acquisition to typically 5 s and to synchronise it with the focal plane crossing of a very bright star. We use a customised version of the \emph{Gaia} observing schedule tool (GOST) to perform the time prediction. The process is typically run twice a year and uses the proper-motion corrected coordinates of the 230 $G<3$ stars as input. The output is written to an interface file that contains the focal plane crossing time (accurate to $\sim$0.5 s), the CCD row, and the field of view for every event. 

\subsection{Commands sent to the Gaia spacecraft}
The interface file with the crossing times is taken as input to a software module called the Payload Operation System (POS) of the Science Operations Centre (SOC) that produces a Payload Operation Request (POR) for every event. The input file is first processed by the scheduler module of the POS, which automatically checks for conflicts amongst the SIF commanding itself and with any other commanding on the payload module, essentially eliminating bright star observations that overlap with other commands or violate the commanding rules. The scheduler also deletes bright star commanding that interferes with other on-board activities like the routinely executed in-orbit control maneuvers or calibration activities.

Then, the POS generates one POR per bright star SIF observation. A POR consists of a file in XML format containing all the telecommands to be executed on-board and their execution times. In this case, this is to acquire a full-frame SkyMapper image at the predicted star-crossing time by using the Service Interface Function\cite{LL:GAIA.ASU.ICD.PLM.00012} (SIF). Finally, the collection of PORs is stored in the Configuration Data Base and sent to the Mission Operations Center (MOC), where the PORs are validated and up-loaded to the \emph{Gaia} spacecraft\cite{Martin-Fleitas:2016}.

\begin{figure}[h!]\begin{center}
\includegraphics[width = \linewidth,trim=0cm -5mm 0 0cm, clip]{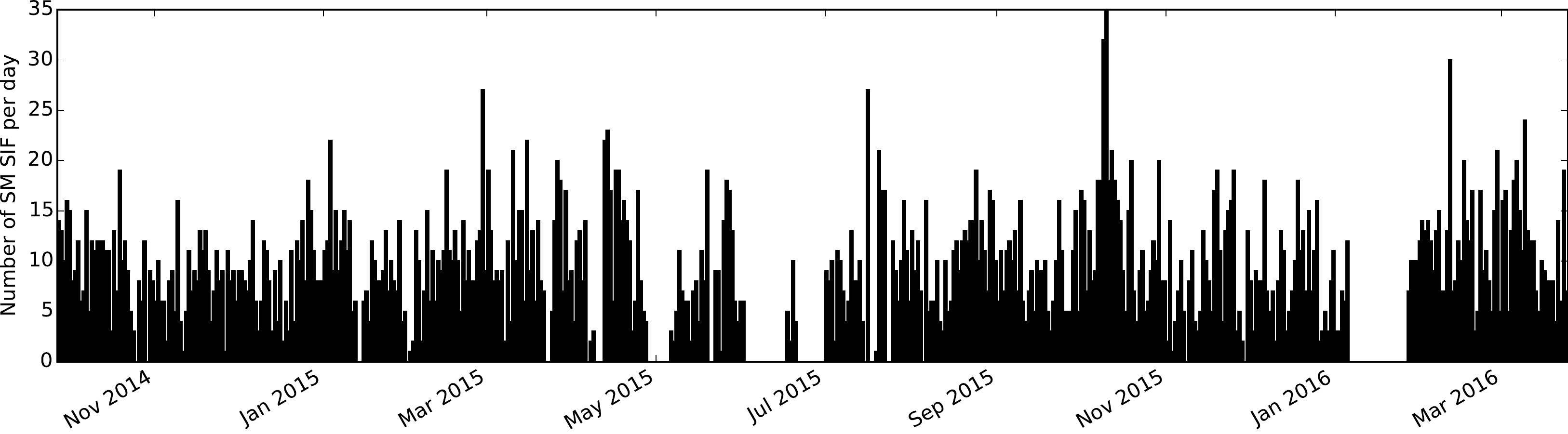} 
\caption{Number of SkyMapper images acquired per day as a function of time. The variation in number per day reflects the distribution of bright stars in the sky and the \emph{Gaia} scan law, whereas the gaps around June 2015 and January 2016 were caused by operational constraints.}
\label{fig:SIFperday}\end{center}\end{figure}

\subsection{Data retrieval and initial processing}
The Service Interface Function is a protocol able to read areas of the Video Processing Unit (VPU) memory, particularly memory buffers, and store them on board in data packets called `SIF packets`. The SIF telecommands are sent to the VPU via a dedicated MIL-STD-1553B link from the Command and Data Management Unit (CDMU). The SIF packets are sent from the VPU through the SpaceWire link to the Payload Data Handling Unit (PDHU), where it is stored until it is downlinked to ground like all the other nominally generated telemetry. A SIF sample is a sequence of bytes characterised by the memory address of the first byte, length, and an on-board acquisition time. SIF samples are grouped in SIF frames and stored in the Payload Data Handling Unit as SIF packets. SkyMapper raw data is stored locally in rolling buffers called VPU Raw Data Buffers (one per SkyMapper). SIF samples are read with the corresponding buffer memory base address, offset, sample size and sample step in sync with the reading of pixels at the output amplifier of the SkyMapper CCD when the target star is crossing the focal plane. This makes it possible to retrieve the SkyMapper raw samples for a star before they are overwritten. The telemetry generated for a bright star observation of 5-s duration is 8 MBytes.

The SIF packets containing the raw CCD data are downlinked from the spacecraft with high-priority and then persisted to data storage by the nominal data-processing pipeline\cite{Siddiqui:2014aa}. In order to extract the CCD data from the SIF packets and reconstruct the full-frame SkyMapper images for each acquisition, a reconstruction tool is periodically run over the latest SIF data\cite{LL:CMC-010}. This produces a repository of raw CCD observations for each bright star transit in a convenient format (e.g.\ FITS) ready for analysis.

\subsection{Current status}
SkyMapper full-frame images of very bright stars are acquired routinely since the beginning of the nominal \emph{Gaia} mission in September 2014. Here we present the inventory of images acquired over 1.5 years (547 days) between 2014-09-25 and 2016-03-25. A total of 4672 images were obtained at an average rate of 9.9 images per day. For every star in Table \ref{tab:vbsstars}, we have obtained between 6 and 69 images and the average number of images per star is 20, where the large variation stems from the \emph{Gaia} scan law which results in a variable sampling of different sky-regions\cite{de-Bruijne:2012kx}. Figure \ref{fig:SIFperday} shows the number of images that were acquired each day. 

\subsection{Extracting photocentres of $G<3$ stars from SkyMapper images}
Our goal is to extract precise astrometric measurements from the SkyMapper images and to make them available to the community in a similar fashion as the nominal \emph{Gaia} data products. However, this special observing mode was not intended to produce scientific data and therefore there were no pre-existing tools for data reduction and analysis. Consequently, there is no pipeline or any other software that is ready to process these data.
We are developing an offline prototype data reduction scheme that will allow us to go from raw SkyMapper images to calibrated \emph{Gaia} astrometry for the 230 observed stars. This is ongoing at the time of writing and we describe the current status.

\begin{figure}[h!]
\begin{center}
\includegraphics[width = 0.8\linewidth]{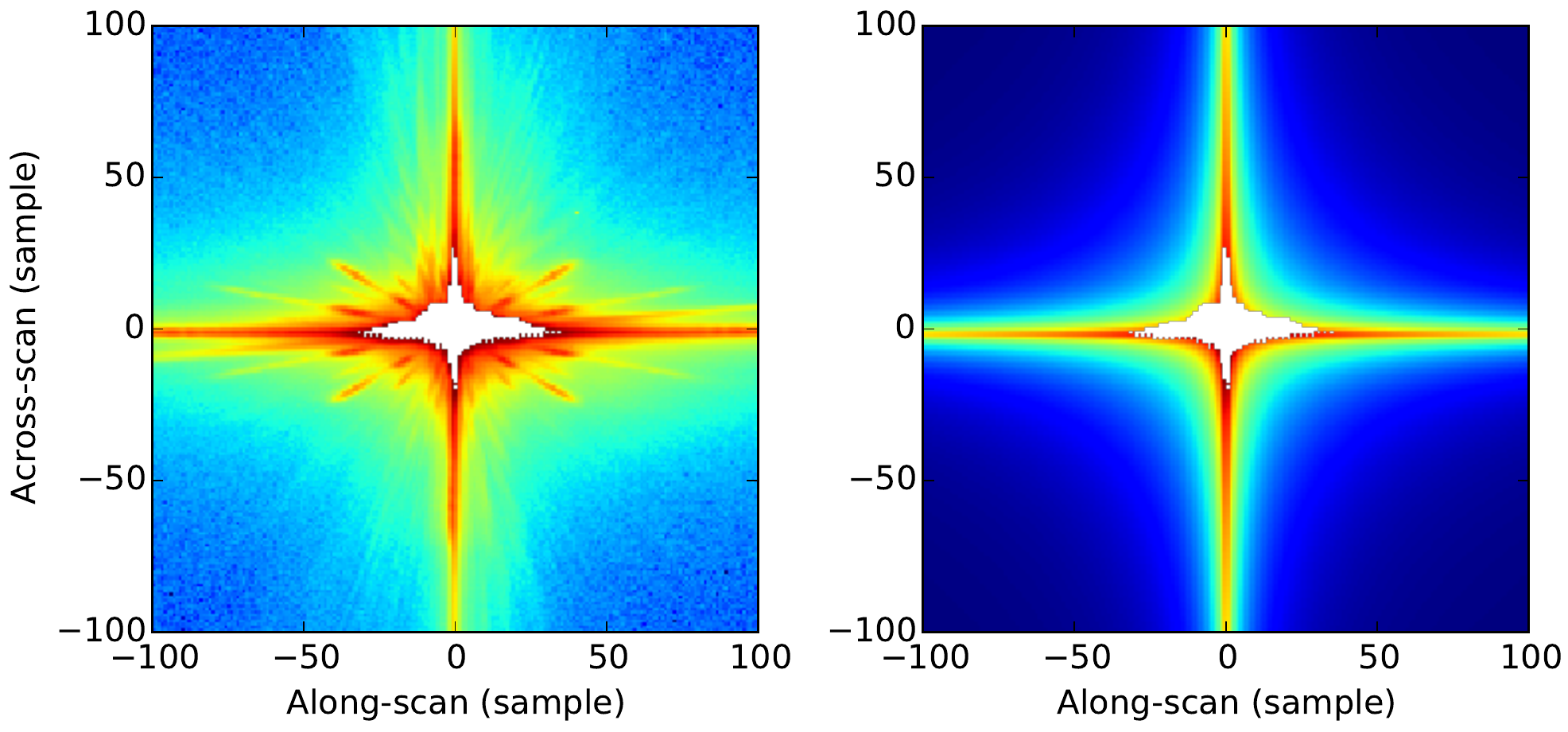}\vspace{2mm}
\caption{\emph{Left:} 200$\times$200 sample cut-out of a SkyMapper full-frame image acquired during the nominal mission. AC is vertical and AL is horizontal and the image size is $70\farcs6 \times 23\farcs5$ on the sky. \emph{Right:} The corresponding 2-dimensional representation of a theoretical \emph{Gaia} PSF is shown on the right. White areas indicate saturated pixels that are masked during the analysis. Both images are shown on a logarithmic stretch.}
\label{fig:fitExample}\end{center}\end{figure}

The basic idea is to adapt the standard \emph{Gaia} data analysis principles to the specific format and content of the data. A \emph{Gaia} astrometric measurement consists of a pair of observables given as observing time, i.e.\ the along-scan coordinate (AL), and across-scan (AC) location on a CCD. These are in principle accessible in the same way for nominally obtained data and for the images taken with the special mode, however there are some important differences: nominal data of bright stars in the astrometric field are acquired from unbinned pixels in a 18x12 pixel window with an adaptive integration time that can be as short as 16 ms (gate 4). The SkyMapper images are acquired as samples that consist of a binned 2$\times$2 pixel areas, which cover a full CCD for a typical duration of 5 s (corresponding to a pixel area of 5088$\times$1966 per image, i.e.\ 2544$\times$983 samples), and the integration time is fixed to 2.85 s (gate 12). The longer integration time results in stronger saturation, but the much larger recorded image area allows us to recover the astrometric information. 

To obtain a preliminary and coarse photocentre location, we assumed that the standard theoretical \emph{Gaia} PSF model, which is the product of two one-dimensional Line Spread Functions (LSFs)\cite{LL:LL-088}, can be applied to the SkyMapper images. Figure \ref{fig:fitExample} shows the comparison between an example of a real image (left) and the model PSF, which has been adjusted for across-scan and along-scan position and amplitude, where saturated pixels have been masked. Whereas there is general agreement between data and model, some features like the radial fine structure close to the PSF core are not readily reproduced by this simple model. We are working on improving the PSF modelling strategy. Whereas the comprehensive PSF modelling of SkyMapper data for very bright stars poses major challenges, it is important to realise that the use of SkyMapper data allows us to measure the location of the PSF wings and diffraction pattern out to distances of several hundred pixels from the PSF core, which will facilitate the determination of the photocentre. 

\begin{figure}[h!]\begin{center}
\includegraphics[width = 0.60\linewidth,trim=0cm 0cm 0 0cm, clip]{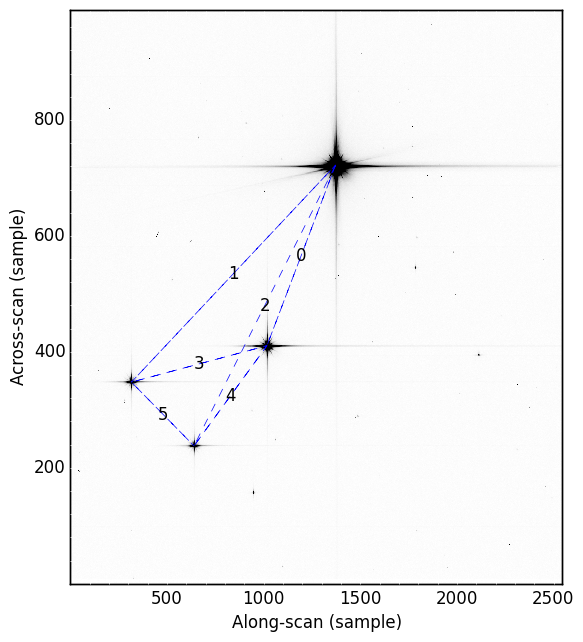} 
\caption{Example of a raw SkyMapper full-frame image covering $5\farcm8 \times 4\farcm9$ acquired during the nominal mission (logarithmic stretch). The very bright star is located in the upper right. Three additional bright stars and many faint stars can be seen as well. There are six relative astrometric measurements labelled 0--5 that we use for local validation.}
\label{fig:SIFexample}\end{center}\end{figure}

To validate the results of our PSF fitting, we adopted a two-stage approach. In the first step, we perform a `local` validation, where we use relative in-frame astrometry measured within one image (see Fig.\ \ref{fig:SIFexample}) to establish the consistency between several independent observations of the same objects and with external astrometric catalogues. In a second step, we perform a global validation by using a future \emph{Gaia} global astrometric solution to analyse the accuracy of individual astrometric measurements obtained from SkyMapper images.

\section{GETTING MORE OUT OF $G<2$ STARS}\label{sec:vo}
The SkyMapper imaging technique discussed in the previous section allows us to record very bright star data with one CCD only.  For $G<2$ stars, where natural detection by the onboard algorithms declines sharply, no other CCD in the \emph{Gaia} focal plane acquires data, although they would be valuable for astrometry, photometry, and spectroscopy. We are testing an observing technique that allows us to overcome this limitation as well. It is termed `virtual object synchronisation` and was briefly introduced in Ref.\ \citenum{Martin-Fleitas:2014aa}, where we also reported on preliminary tests undertaken during commissioning.

\subsection{Principles of virtual object synchronisation}
In addition to the normal observing windows that are being assigned to stellar objects in real-time, the \emph{Gaia} on-board software allows for the insertion of an additional set of externally defined objects with associated windows. Those are called Virtual Objects (VO) and are intended to force the observation of a pattern of windows placed at known locations. They usually fall on `empty` regions of the sky and, for instance, serve to estimate the sky background in the data reduction pipeline of the ground-segment. The idea of VO synchronised observations is to predict the focal plane crossing of a very bright star and to place a VO window on top of it. This is possible while preserving the original VO pattern because this pattern does not fully exhaust the VO resource budget available, and a limited number of additional virtual objects can be appended to it.

\begin{figure}[h!]\begin{center}
\includegraphics[height = 0.27\linewidth,trim=0 0cm 0 0cm, clip]{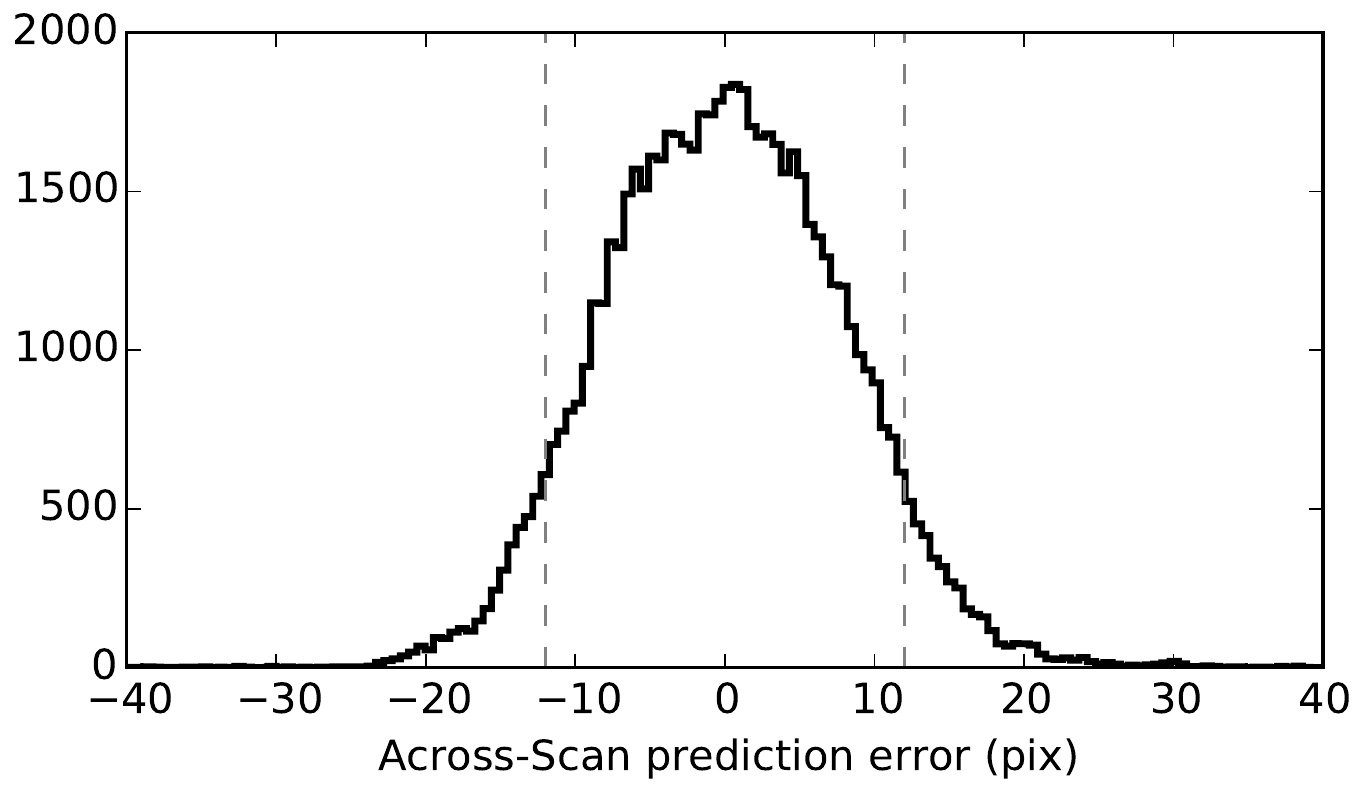} 
\includegraphics[height = 0.25\linewidth,trim=0 -2cm 0 0cm, clip, angle=0]{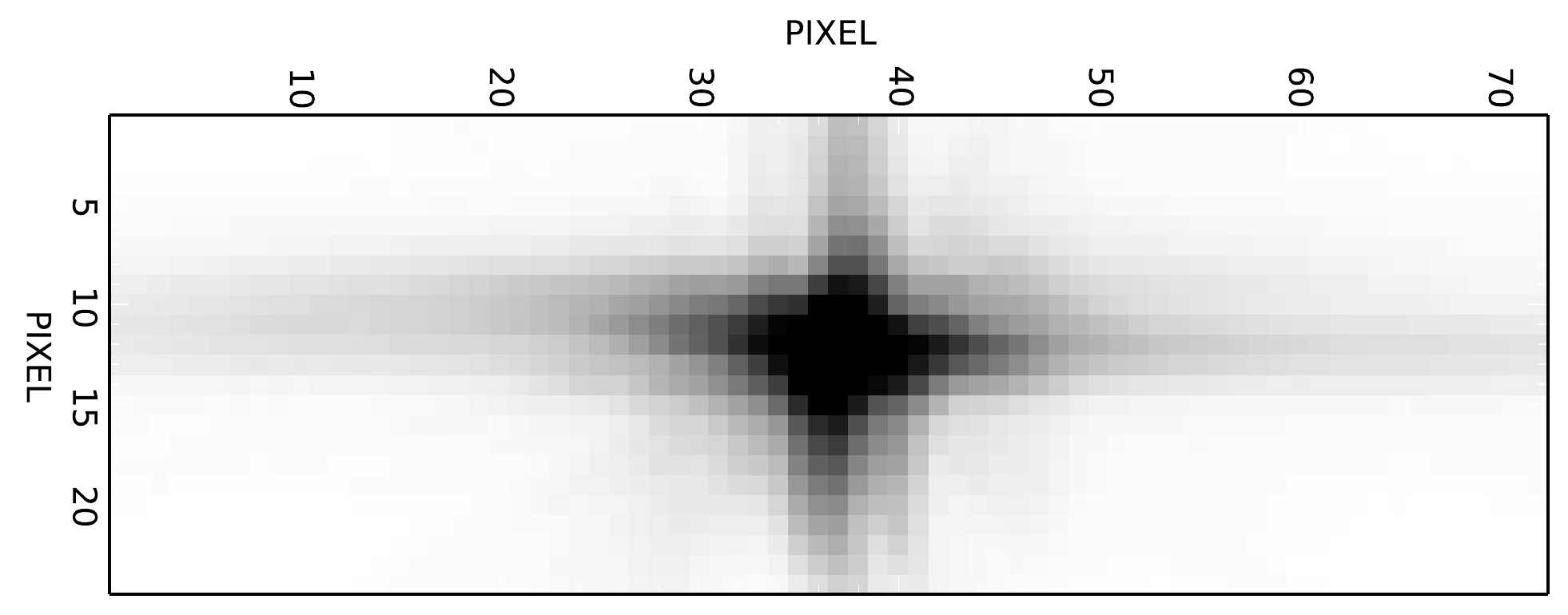}

\caption{\emph{Left:} Histogram of prediction errors in across-scan coordinate for the events discussed in Sect.\ \ref{sec:brightLim}. Dashed lines indicate the $\pm$12 pixel extent of the composite VO windows we use. \emph{Right}: Composite of 4$\times$2 VO windows (total of 72$\times$24 pixels) that we used to capture the image of a $G=2.1$ star with the astrometric field CCD AF2. Along-scan is horizontal and across-scan is vertical  in this image.}
\label{fig:prederror}\end{center}\end{figure}

Because the largest available VO window has the same size as a bright-object window (called class0, which is $18$ AL $\times$ 12 AC pixels), VO synchronisation relies on highly accurate predictions of transit times and across-scan positions in the \emph{Gaia} focal plane. We achieved this by modelling the difference between GOST predictions and actual \emph{Gaia} observations\cite{JSA-006}, which allowed us to demonstrate residual prediction errors (r.m.s.) of $\sim$5 pixel in AL and $\sim$8 pixel in AC (Fig.\ \ref{fig:prederror}). These errors are still considerable compared to the size of one single VO window ($18\times12$ pixels). Therefore we use composite VO windows that are concatenations of several $18\times12$ pixel windows to ensure that the stellar image is captured (see Fig.\ \ref{fig:prederror}). A maximum of 2 class0 VO windows can be commanded at any given time for a single CCD row, i.e.\ the across-scan extent of a composite window is limited to 24 pixels.

\subsection{Successful in-flight demonstration}
To test the VO synchronisation technique during nominal \emph{Gaia} operations, we introduced 93 additional virtual objects in the routine VO pattern over a time range of 3 days in June 2015. We used 36$\times$24 pixel composite windows that were predicted to fall on bright stars with magnitudes in the range $G=1.1-5.0$. The resulting data showed that all stars have been either fully or partially observed. In 34 out of 93 cases ($37 \pm 5$ \%)  the stellar core was missed, but it will be possible to extract along-scan astrometry from the PSF wing captured in the window. In 59 out of 93 cases ($63 \pm 5$ \%), the star core was captured which in principle allows full exploitation of the astrometric, photometric, and spectroscopic data.

To estimate the astrometric precision that can be achieved with such observations, we computed the Cramer Rao lower bound \cite{U.Bastian:2004zr, LL:AMF-010, Mendez:2013aa} for the composite windows acquired during the June 2015 test. For stars where the core was observed, the estimated astrometric precision from a complete focal plane transit is consistently better than 100 $\mu$as. Even for the VOs where the star core was missed, this estimate is better than 1 mas in all but 5 cases\cite{JSA-008}.  The final astrometric error of these observations will be the sum of the formal precision and the residual calibration error, which will therefore be larger than the numbers derived as the ultimate limit from the Cramer Rao lower bound. Here, we do not quantify how well these data can be calibrated. 

\subsection{Outlook}
Virtual object synchronisation is a promising technique to force the observation of extremely bright objects. In addition to astrometry, absolute spectrophotometry of bright spectrophotometric standards would also be made possible. Its feasibility was demonstrated both from the prediction and implementation point of view. Assuming a scenario where the 50 brightest stars ($G\lesssim1.75$) are observed with virtual object synchronisation, the impact on the mission's dead time, the \emph{Gaia} telemetry budget, and the required resources were estimated and found to be minor\cite{LL:AMF-024}.  Given the high success rate of our test observations, we estimate that the \emph{Gaia} mission would benefit from adopting VO synchronisation for the brightest 50 stars in operation.  

\section{OTHER APPLICATIONS OF SkyMapper IMAGING}
\begin{figure}[h!]\begin{center}
\includegraphics[width = 0.9\linewidth,trim=0 0cm 0 0cm, clip]{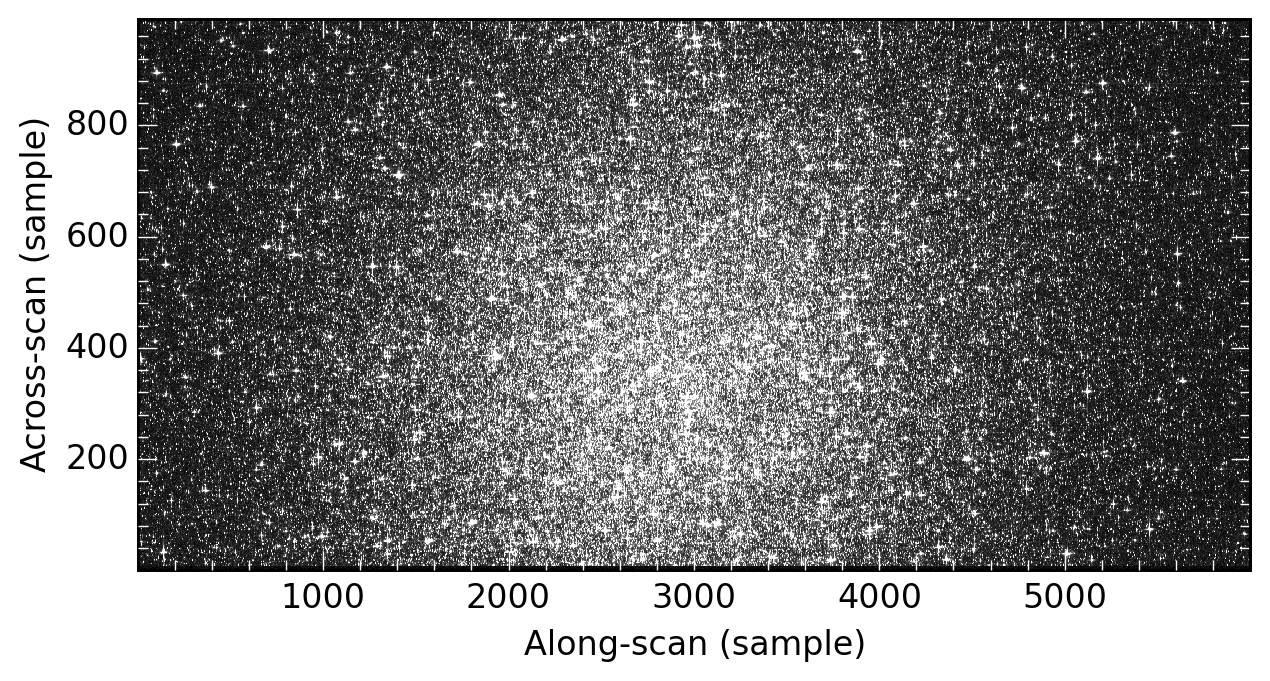} 
\caption{SkyMapper image of the central region of the globular cluster Omega Cen. The image was obtained with a single CCD and an exposure time of 2.85 s (gate 12). It took \emph{Gaia} 11.8 s to sweep across this field that covers 11.8\arcmin$\times$5.8\arcmin. See \url{http://www.cosmos.esa.int/web/gaia/iow_20160504}.}
\label{fig:omegaCen}\end{center}\end{figure}

Whereas the \emph{Gaia} onboard object detection warrants observations completeness for stars down to $G\simeq20$, this faint limit is not reached in the densest regions of the sky due to the limited on-board resources. This adversely affects the \emph{Gaia} observations of the two most prominent dense regions, the globular clusters Omega Cen  and NGC 6522 (Baade's window). To mitigate the effects of crowding on the observation completeness and to support the scientific analysis of these two fields, we are regularly acquiring SkyMapper full-frame images of these regions. An example image of a portion of Omega Cen is shown in Figure \ref{fig:omegaCen}. A preliminary analysis of such images indicates that they can lead to a significant gain in the detection of faint sources in the central part of the cluster, as compared to the natural detections achieved by the nominal on-board algorithms\cite{LL:AMF-023}.

\section{CONCLUSIONS} 
We presented our efforts to make \emph{Gaia} observe the stars brighter than $G=6$, i.e.\ the naked-eye stars, which would not have been included in the \emph{Gaia} survey according to the original mission design. This work is motivated by the scientific potential that \emph{Gaia} data of those stars offer, in particular in the field of stellar physics, extrasolar planets, and binary stars. We presented \emph{Gaia}'s bright star detection performance, which shows that stars fainter then $G=3$ are being naturally observed almost as frequent as stars in the nominal $G>6$ magnitude range. Since the beginning of the \emph{Gaia} science mission, the extremely bright stars $G<3$ are being observed with a special mode that produces full-frame images with the SkyMapper CCD. Table \ref{tab:data} gives an overview of the data collected for very bright stars.

We presented a technical solution that would allow the collection of \emph{Gaia} astrometric and spectrophotometric data for extremely bright stars ($G<2$). It is based on virtual object synchronisation and was successfully demonstrated during nominal \emph{Gaia} operations. It is however not yet clear whether this mode will be adopted for the remainder of the \emph{Gaia} mission. We also showed how the special observing mode with full-frame SkyMapper images that we developed for bright stars may be used to enhance the \emph{Gaia} survey of regions with extremely high stellar densities like Omega Cen and Baade's window.

The current operational scheme is making \emph{Gaia} astrometry effectively complete at the bright end. Whereas preliminary estimates promise an astrometric performance for very bright stars that is unrivaled by any other planned instrument/mission, there is still uncertainty about the achievable astrometric accuracy, which ultimately will be limited by calibration errors. So far, we have mostly concentrated on making sure that very bright star data are collected by \emph{Gaia}, which we achieved as discussed here. Currently, we are addressing the astrometric accuracy of these observations and we will describe our results in forthcoming publications. Eventually, \emph{Gaia} astrometry of very bright stars will be made publicly available as part of a future \emph{Gaia} data release.

\begin{table}[h!]
\caption{Overview of \emph{Gaia} data collected for $G<6$ stars. VOS stands for virtual object synchronisation (Sect.\ \ref{sec:vo}). SIF stands for Service Interface Function, which yields full-frame SkyMapper images (Sect.\ \ref{sec:SIF}).}
\label{tab:data}
\begin{center}
\begin{tabular}{lccc}
\hline
\hline
Datatype    & Acronym              & $G<3$             &$3<G<5.7$ \\
\hline
SkyMapper full frame & SM SIF &    Yes               &No\\
SkyMapper window & SM1/2&No (unless detected or VOS)	& Yes \\
Astrometric field window & AF1..9 &No (unless detected or VOS)	& Yes \\
Photometers & BP/RP & 	No (unless detected or VOS)	& Yes \\
Spectrometer & RVS & If detected	& Yes \\
\hline
\end{tabular}
\end{center}
\end{table}

\acknowledgments
The authors wish to thank the \emph{Gaia} Science Team and Data Processing and Analysis Consortium (DPAC) for their support. We also kindly acknowledge the \emph{Gaia} Data Processing Centre Spain (DPCE), \emph{Gaia} Science Operations (SOC), and Mission Operations Centre (MOC) teams for their continuous support. J.S. is supported by an ESA Research Fellowship in Space Science.\\
This work has made use of results from the ESA space mission Gaia, the data from which were processed by the Gaia Data Processing and Analysis Consortium (DPAC). Funding for the DPAC has been provided by national institutions, in particular the institutions participating in the Gaia Multilateral Agreement. The Gaia mission website is: \href{http://www.cosmos.esa.int/gaia}{http://www.cosmos.esa.int/gaia}.\\
This research made use of the databases at the Centre de Donn\'ees astronomiques de Strasbourg (\url{http://cds.u-strasbg.fr}), NASA's Astrophysics Data System Service (\url{http://adsabs.harvard.edu/abstract\_service.html}), the paper repositories at arXiv, of APLpy, an open-source plotting package for Python hosted at \href{http://aplpy.github.com}{http://aplpy.github.com}, and of Astropy, a community-developed core Python package for Astronomy\cite{Astropy-Collaboration:2013aa}. The authors also made use of SCIPY\cite{Jones:2001aa}, NUMPY\cite{Oliphant2007}, IPYTHON\cite{Perez2007}, and MATPLOTLIB\cite{hunter2007}.

\bibliography{biblio}
\bibliographystyle{spiebib_mod}   

\clearpage
\begin{appendix}
\section{Data tables}

\begin{table*}[h!]
\caption{Parameters for very bright stars with known planets. $G$ is estimated using Refs. \citenum{Jordi:2010kx} and \citenum{de-Bruijne:2014aa}, spectral types are from Ref. \citenum{ESA:1997vn}, star and planet parameters are from \href{http://exoplanets.org}{exoplanets.org}.}
\label{tab:1a} 
\footnotesize
\centering  
\begin{tabular}{c r r r r r r r r} 
\hline\hline 
Nr.        &Name& $V$   &$G$ & Sp. type & $P$  & $d$ &$a_{1,min}$\\
             &         & (mag)& (mag) &               & (day)& (pc) & ($\mu$as) \\  
\hline 
1 & HD 39091 b & 5.7 & 5.5 & G3IV  & 2151.0 & 18.3 & 1628.2\\
 2 & epsilon Eri b & 3.7 & 3.5 & K2V  & 2500.0 & 3.2 & 1286.8\\
 3 & upsilon And d & 4.1 & 4.0 & F8V  & 1278.1 & 13.5 & 559.5\\
 4 & HD 190360 b & 5.7 & 5.5 & G6IV+...  & 2915.0 & 15.9 & 372.9\\
 5 & HD 33564 b & 5.1 & 4.9 & F6V  & 388.0 & 20.9 & 372.5\\
 6 & 47 UMa b & 5.0 & 4.9 & G0V  & 1078.0 & 14.1 & 340.8\\
 7 & iota Dra b & 3.3 & 3.0 & K2III  & 511.1 & 31.0 & 326.8\\
 8 & beta Gem b & 1.1 & 0.9 & K0IIIvar  & 589.6 & 10.4 & 214.5\\
 9 & gamma Leo A b & 2.1 & 1.7 & K0III  & 428.5 & 39.9 & 203.5\\
10 & 70 Vir b & 5.0 & 4.8 & G5V  & 116.7 & 18.0 & 172.8\\
11 & gamma Cep b & 3.2 & 2.9 & K1IV  & 905.6 & 14.1 & 161.1\\
12 & 7 CMa b & 4.0 & 3.7 & K1III+...  & 763.0 & 19.8 & 157.9\\
13 & bet UMi b & 2.1 & 1.6 & K4IIIvar  & 522.3 & 40.1 & 146.7\\
14 & 18 Del b & 5.5 & 5.3 & G6III:  & 993.3 & 75.3 & 144.3\\
15 & mu Ara b & 5.1 & 5.0 & G5V  & 643.2 & 15.5 & 142.8\\
16 & HD 139357 b & 6.0 & 5.6 & K4III:  & 1125.7 & 118.1 & 140.6\\
17 & 47 UMa c & 5.0 & 4.9 & G0V  & 2391.0 & 14.1 & 124.4\\
18 & HD 89744 b & 5.7 & 5.6 & F7V  & 256.8 & 39.4 & 120.2\\
19 & epsilon Tau b & 3.5 & 3.2 & K0III  & 594.9 & 45.0 & 115.3\\
20 & HD 147513 b & 5.4 & 5.2 & G3/G5V  & 528.4 & 12.8 & 107.5\\
21 & kappa CrB b & 4.8 & 4.5 & K0III-IV  & 1300.0 & 30.5 & 106.1\\
22 & HD 4732 c & 5.9 & 5.7 & K0III  & 2732.0 & 57.9 & 103.0\\
23 & HD 60532 c & 4.5 & 4.3 & F6V  & 604.0 & 25.3 & 101.8\\
24 & 11 Com b & 4.8 & 4.4 & G8III  & 326.0 & 88.9 & 99.3\\
25 & HD 11977 b & 4.7 & 4.4 & G5III  & 711.0 & 67.1 & 93.8\\
26 & HD 10647 b & 5.5 & 5.4 & F8V  & 1003.0 & 17.4 & 93.4\\
27 & eta Cet c & 3.5 & 3.1 & K2III  & 744.5 & 38.0 & 93.1\\
28 & iota Hor b & 5.4 & 5.3 & G3IV  & 302.8 & 17.2 & 91.7\\
29 & omicron UMa b & 3.4 & 3.1 & G4II-III  & 1630.0 & 54.9 & 90.9\\
30 & HD 30562 b & 5.8 & 5.6 & F8V  & 1157.0 & 26.4 & 88.2\\
31 & upsilon And c & 4.1 & 4.0 & F8V  & 241.3 & 13.5 & 85.9\\
32 & bet Cnc b & 3.5 & 3.0 & K4III  & 605.2 & 93.0 & 78.5\\
33 & HD 120084 b & 5.9 & 5.6 & G7III:  & 2082.0 & 97.7 & 78.0\\
34 & 11 UMi b & 5.0 & 4.6 & K4III  & 516.2 & 122.1 & 73.5\\
35 & epsilon CrB b & 4.1 & 3.8 & K3III  & 417.9 & 67.9 & 72.8\\
36 & epsilon Ret b & 4.4 & 4.1 & K2IV  & 428.1 & 18.2 & 69.3\\
37 & alpha Ari b & 2.0 & 1.7 & K2III  & 380.0 & 20.2 & 68.9\\
38 & 4 UMa b & 5.8 & 4.2 & K2III  & 269.3 & 78.5 & 61.3\\
39 & HD 19994 b & 5.1 & 4.9 & F8V  & 466.2 & 22.6 & 53.6\\
40 & 81 Cet b & 5.7 & 5.4 & G5III:  & 952.7 & 100.9 & 53.3\\
 \hline
\end{tabular} 
\end{table*}

\begin{table}[h!]
\caption{\emph{Hipparcos} identifiers (with hyperlink to Simbad database), $V$ magnitudes, and $G$ magnitudes of the 230 stars that are being observed with SkyMapper images (170 stars are listed here, the remaining stars in Table \ref{tab:vbsstars2}). $G$ was computed from $V$ and $V-I$ using the relations given in Ref.\ \citenum{Jordi:2010kx}.}\vspace{0mm}
\label{tab:vbsstars}
\begin{center}
\begin{tabular}{rrr|rrr|rrr|rrr}
\hline
\hline
HIP & $V$ & $G$ & HIP & $V$ & $G$ & HIP & $V$ & $G$ & HIP & $V$ & $G$ \\
   & (mag) & (mag) &    & (mag) & (mag) &    & (mag) & (mag) &    & (mag) & (mag) \\ 
\hline
\href{http://simbad.u-strasbg.fr/simbad/sim-basic?Ident=HIP000677&submit=SIMBAD+search}{677} & 2.1 & 2.1 & \href{http://simbad.u-strasbg.fr/simbad/sim-basic?Ident=HIP025606&submit=SIMBAD+search}{25606} & 2.8 & 2.6 & \href{http://simbad.u-strasbg.fr/simbad/sim-basic?Ident=HIP045860&submit=SIMBAD+search}{45860} & 3.1 & 2.6 & \href{http://simbad.u-strasbg.fr/simbad/sim-basic?Ident=HIP067927&submit=SIMBAD+search}{67927} & 2.7 & 2.5 \\
\href{http://simbad.u-strasbg.fr/simbad/sim-basic?Ident=HIP000746&submit=SIMBAD+search}{746} & 2.3 & 2.2 & \href{http://simbad.u-strasbg.fr/simbad/sim-basic?Ident=HIP025930&submit=SIMBAD+search}{25930} & 2.2 & 2.2 & \href{http://simbad.u-strasbg.fr/simbad/sim-basic?Ident=HIP045941&submit=SIMBAD+search}{45941} & 2.5 & 2.5 & \href{http://simbad.u-strasbg.fr/simbad/sim-basic?Ident=HIP068002&submit=SIMBAD+search}{68002} & 2.5 & 2.5 \\
\href{http://simbad.u-strasbg.fr/simbad/sim-basic?Ident=HIP001067&submit=SIMBAD+search}{1067} & 2.8 & 2.8 & \href{http://simbad.u-strasbg.fr/simbad/sim-basic?Ident=HIP025985&submit=SIMBAD+search}{25985} & 2.6 & 2.5 & \href{http://simbad.u-strasbg.fr/simbad/sim-basic?Ident=HIP046390&submit=SIMBAD+search}{46390} & 2.0 & 1.5 & \href{http://simbad.u-strasbg.fr/simbad/sim-basic?Ident=HIP068702&submit=SIMBAD+search}{68702} & 0.6 & 0.6 \\
\href{http://simbad.u-strasbg.fr/simbad/sim-basic?Ident=HIP002021&submit=SIMBAD+search}{2021} & 2.8 & 2.7 & \href{http://simbad.u-strasbg.fr/simbad/sim-basic?Ident=HIP026241&submit=SIMBAD+search}{26241} & 2.8 & 2.7 & \href{http://simbad.u-strasbg.fr/simbad/sim-basic?Ident=HIP046701&submit=SIMBAD+search}{46701} & 3.2 & 2.6 & \href{http://simbad.u-strasbg.fr/simbad/sim-basic?Ident=HIP068895&submit=SIMBAD+search}{68895} & 3.2 & 2.9 \\
\href{http://simbad.u-strasbg.fr/simbad/sim-basic?Ident=HIP002081&submit=SIMBAD+search}{2081} & 2.4 & 2.1 & \href{http://simbad.u-strasbg.fr/simbad/sim-basic?Ident=HIP026311&submit=SIMBAD+search}{26311} & 1.7 & 1.7 & \href{http://simbad.u-strasbg.fr/simbad/sim-basic?Ident=HIP047908&submit=SIMBAD+search}{47908} & 3.0 & 2.8 & \href{http://simbad.u-strasbg.fr/simbad/sim-basic?Ident=HIP068933&submit=SIMBAD+search}{68933} & 2.1 & 1.8 \\
\href{http://simbad.u-strasbg.fr/simbad/sim-basic?Ident=HIP003092&submit=SIMBAD+search}{3092} & 3.3 & 2.9 & \href{http://simbad.u-strasbg.fr/simbad/sim-basic?Ident=HIP026451&submit=SIMBAD+search}{26451} & 3.0 & 3.0 & \href{http://simbad.u-strasbg.fr/simbad/sim-basic?Ident=HIP048002&submit=SIMBAD+search}{48002} & 2.9 & 2.8 & \href{http://simbad.u-strasbg.fr/simbad/sim-basic?Ident=HIP069673&submit=SIMBAD+search}{69673} & -0.1 & -0.4 \\
\href{http://simbad.u-strasbg.fr/simbad/sim-basic?Ident=HIP003179&submit=SIMBAD+search}{3179} & 2.2 & 1.9 & \href{http://simbad.u-strasbg.fr/simbad/sim-basic?Ident=HIP026634&submit=SIMBAD+search}{26634} & 2.6 & 2.6 & \href{http://simbad.u-strasbg.fr/simbad/sim-basic?Ident=HIP049669&submit=SIMBAD+search}{49669} & 1.4 & 1.3 & \href{http://simbad.u-strasbg.fr/simbad/sim-basic?Ident=HIP071075&submit=SIMBAD+search}{71075} & 3.0 & 3.0 \\
\href{http://simbad.u-strasbg.fr/simbad/sim-basic?Ident=HIP003419&submit=SIMBAD+search}{3419} & 2.0 & 1.8 & \href{http://simbad.u-strasbg.fr/simbad/sim-basic?Ident=HIP026727&submit=SIMBAD+search}{26727} & 1.7 & 1.7 & \href{http://simbad.u-strasbg.fr/simbad/sim-basic?Ident=HIP050371&submit=SIMBAD+search}{50371} & 3.4 & 2.9 & \href{http://simbad.u-strasbg.fr/simbad/sim-basic?Ident=HIP071352&submit=SIMBAD+search}{71352} & 2.3 & 2.3 \\
\href{http://simbad.u-strasbg.fr/simbad/sim-basic?Ident=HIP004427&submit=SIMBAD+search}{4427} & 2.1 & 2.1 & \href{http://simbad.u-strasbg.fr/simbad/sim-basic?Ident=HIP027366&submit=SIMBAD+search}{27366} & 2.1 & 2.1 & \href{http://simbad.u-strasbg.fr/simbad/sim-basic?Ident=HIP050583&submit=SIMBAD+search}{50583} & 2.0 & 1.7 & \href{http://simbad.u-strasbg.fr/simbad/sim-basic?Ident=HIP071681&submit=SIMBAD+search}{71681} & 1.4 & 1.1 \\
\href{http://simbad.u-strasbg.fr/simbad/sim-basic?Ident=HIP005447&submit=SIMBAD+search}{5447} & 2.1 & 1.4 & \href{http://simbad.u-strasbg.fr/simbad/sim-basic?Ident=HIP027628&submit=SIMBAD+search}{27628} & 3.1 & 2.8 & \href{http://simbad.u-strasbg.fr/simbad/sim-basic?Ident=HIP050801&submit=SIMBAD+search}{50801} & 3.1 & 2.4 & \href{http://simbad.u-strasbg.fr/simbad/sim-basic?Ident=HIP071683&submit=SIMBAD+search}{71683} & -0.0 & -0.2 \\
\href{http://simbad.u-strasbg.fr/simbad/sim-basic?Ident=HIP006686&submit=SIMBAD+search}{6686} & 2.7 & 2.6 & \href{http://simbad.u-strasbg.fr/simbad/sim-basic?Ident=HIP027989&submit=SIMBAD+search}{27989} & 0.5 & -0.6 & \href{http://simbad.u-strasbg.fr/simbad/sim-basic?Ident=HIP052419&submit=SIMBAD+search}{52419} & 2.7 & 2.7 & \href{http://simbad.u-strasbg.fr/simbad/sim-basic?Ident=HIP071860&submit=SIMBAD+search}{71860} & 2.3 & 2.3 \\
\href{http://simbad.u-strasbg.fr/simbad/sim-basic?Ident=HIP006867&submit=SIMBAD+search}{6867} & 3.4 & 2.8 & \href{http://simbad.u-strasbg.fr/simbad/sim-basic?Ident=HIP028360&submit=SIMBAD+search}{28360} & 1.9 & 1.9 & \href{http://simbad.u-strasbg.fr/simbad/sim-basic?Ident=HIP052727&submit=SIMBAD+search}{52727} & 2.7 & 2.5 & \href{http://simbad.u-strasbg.fr/simbad/sim-basic?Ident=HIP072105&submit=SIMBAD+search}{72105} & 2.4 & 2.1 \\
\href{http://simbad.u-strasbg.fr/simbad/sim-basic?Ident=HIP007588&submit=SIMBAD+search}{7588} & 0.5 & 0.4 & \href{http://simbad.u-strasbg.fr/simbad/sim-basic?Ident=HIP028380&submit=SIMBAD+search}{28380} & 2.6 & 2.6 & \href{http://simbad.u-strasbg.fr/simbad/sim-basic?Ident=HIP052943&submit=SIMBAD+search}{52943} & 3.1 & 2.7 & \href{http://simbad.u-strasbg.fr/simbad/sim-basic?Ident=HIP072607&submit=SIMBAD+search}{72607} & 2.1 & 1.6 \\
\href{http://simbad.u-strasbg.fr/simbad/sim-basic?Ident=HIP008903&submit=SIMBAD+search}{8903} & 2.6 & 2.6 & \href{http://simbad.u-strasbg.fr/simbad/sim-basic?Ident=HIP029655&submit=SIMBAD+search}{29655} & 3.3 & 2.0 & \href{http://simbad.u-strasbg.fr/simbad/sim-basic?Ident=HIP053910&submit=SIMBAD+search}{53910} & 2.3 & 2.3 & \href{http://simbad.u-strasbg.fr/simbad/sim-basic?Ident=HIP072622&submit=SIMBAD+search}{72622} & 2.8 & 2.7 \\
\href{http://simbad.u-strasbg.fr/simbad/sim-basic?Ident=HIP009236&submit=SIMBAD+search}{9236} & 2.9 & 2.8 & \href{http://simbad.u-strasbg.fr/simbad/sim-basic?Ident=HIP030324&submit=SIMBAD+search}{30324} & 2.0 & 2.0 & \href{http://simbad.u-strasbg.fr/simbad/sim-basic?Ident=HIP054061&submit=SIMBAD+search}{54061} & 1.8 & 1.5 & \href{http://simbad.u-strasbg.fr/simbad/sim-basic?Ident=HIP073273&submit=SIMBAD+search}{73273} & 2.7 & 2.7 \\
\href{http://simbad.u-strasbg.fr/simbad/sim-basic?Ident=HIP009640&submit=SIMBAD+search}{9640} & 2.1 & 1.7 & \href{http://simbad.u-strasbg.fr/simbad/sim-basic?Ident=HIP030343&submit=SIMBAD+search}{30343} & 2.9 & 1.9 & \href{http://simbad.u-strasbg.fr/simbad/sim-basic?Ident=HIP054539&submit=SIMBAD+search}{54539} & 3.0 & 2.7 & \href{http://simbad.u-strasbg.fr/simbad/sim-basic?Ident=HIP073714&submit=SIMBAD+search}{73714} & 3.2 & 2.3 \\
\href{http://simbad.u-strasbg.fr/simbad/sim-basic?Ident=HIP009884&submit=SIMBAD+search}{9884} & 2.0 & 1.7 & \href{http://simbad.u-strasbg.fr/simbad/sim-basic?Ident=HIP030438&submit=SIMBAD+search}{30438} & -0.6 & -0.7 & \href{http://simbad.u-strasbg.fr/simbad/sim-basic?Ident=HIP054872&submit=SIMBAD+search}{54872} & 2.6 & 2.5 & \href{http://simbad.u-strasbg.fr/simbad/sim-basic?Ident=HIP074785&submit=SIMBAD+search}{74785} & 2.6 & 2.6 \\
\href{http://simbad.u-strasbg.fr/simbad/sim-basic?Ident=HIP010064&submit=SIMBAD+search}{10064} & 3.0 & 3.0 & \href{http://simbad.u-strasbg.fr/simbad/sim-basic?Ident=HIP031681&submit=SIMBAD+search}{31681} & 1.9 & 1.9 & \href{http://simbad.u-strasbg.fr/simbad/sim-basic?Ident=HIP057632&submit=SIMBAD+search}{57632} & 2.1 & 2.1 & \href{http://simbad.u-strasbg.fr/simbad/sim-basic?Ident=HIP074946&submit=SIMBAD+search}{74946} & 2.9 & 2.8 \\
\href{http://simbad.u-strasbg.fr/simbad/sim-basic?Ident=HIP010826&submit=SIMBAD+search}{10826} & 6.5 & 2.3 & \href{http://simbad.u-strasbg.fr/simbad/sim-basic?Ident=HIP032246&submit=SIMBAD+search}{32246} & 3.1 & 2.7 & \href{http://simbad.u-strasbg.fr/simbad/sim-basic?Ident=HIP058001&submit=SIMBAD+search}{58001} & 2.4 & 2.4 & \href{http://simbad.u-strasbg.fr/simbad/sim-basic?Ident=HIP075097&submit=SIMBAD+search}{75097} & 3.0 & 3.0 \\
\href{http://simbad.u-strasbg.fr/simbad/sim-basic?Ident=HIP011767&submit=SIMBAD+search}{11767} & 2.0 & 1.8 & \href{http://simbad.u-strasbg.fr/simbad/sim-basic?Ident=HIP032349&submit=SIMBAD+search}{32349} & -1.4 & -1.5 & \href{http://simbad.u-strasbg.fr/simbad/sim-basic?Ident=HIP059196&submit=SIMBAD+search}{59196} & 2.6 & 2.6 & \href{http://simbad.u-strasbg.fr/simbad/sim-basic?Ident=HIP075458&submit=SIMBAD+search}{75458} & 3.3 & 3.0 \\
\href{http://simbad.u-strasbg.fr/simbad/sim-basic?Ident=HIP013847&submit=SIMBAD+search}{13847} & 2.9 & 2.8 & \href{http://simbad.u-strasbg.fr/simbad/sim-basic?Ident=HIP032768&submit=SIMBAD+search}{32768} & 2.9 & 2.6 & \href{http://simbad.u-strasbg.fr/simbad/sim-basic?Ident=HIP059316&submit=SIMBAD+search}{59316} & 3.0 & 2.7 & \href{http://simbad.u-strasbg.fr/simbad/sim-basic?Ident=HIP076267&submit=SIMBAD+search}{76267} & 2.2 & 2.2 \\
\href{http://simbad.u-strasbg.fr/simbad/sim-basic?Ident=HIP014135&submit=SIMBAD+search}{14135} & 2.5 & 1.8 & \href{http://simbad.u-strasbg.fr/simbad/sim-basic?Ident=HIP033579&submit=SIMBAD+search}{33579} & 1.5 & 1.5 & \href{http://simbad.u-strasbg.fr/simbad/sim-basic?Ident=HIP059747&submit=SIMBAD+search}{59747} & 2.8 & 2.8 & \href{http://simbad.u-strasbg.fr/simbad/sim-basic?Ident=HIP076297&submit=SIMBAD+search}{76297} & 2.8 & 2.8 \\
\href{http://simbad.u-strasbg.fr/simbad/sim-basic?Ident=HIP014328&submit=SIMBAD+search}{14328} & 2.9 & 2.7 & \href{http://simbad.u-strasbg.fr/simbad/sim-basic?Ident=HIP033856&submit=SIMBAD+search}{33856} & 3.5 & 2.8 & \href{http://simbad.u-strasbg.fr/simbad/sim-basic?Ident=HIP059803&submit=SIMBAD+search}{59803} & 2.6 & 2.6 & \href{http://simbad.u-strasbg.fr/simbad/sim-basic?Ident=HIP077070&submit=SIMBAD+search}{77070} & 2.6 & 2.3 \\
\href{http://simbad.u-strasbg.fr/simbad/sim-basic?Ident=HIP014354&submit=SIMBAD+search}{14354} & 3.3 & 2.0 & \href{http://simbad.u-strasbg.fr/simbad/sim-basic?Ident=HIP033977&submit=SIMBAD+search}{33977} & 3.0 & 3.0 & \href{http://simbad.u-strasbg.fr/simbad/sim-basic?Ident=HIP059929&submit=SIMBAD+search}{59929} & 4.1 & 2.7 & \href{http://simbad.u-strasbg.fr/simbad/sim-basic?Ident=HIP077952&submit=SIMBAD+search}{77952} & 2.8 & 2.8 \\
\href{http://simbad.u-strasbg.fr/simbad/sim-basic?Ident=HIP014576&submit=SIMBAD+search}{14576} & 2.1 & 2.1 & \href{http://simbad.u-strasbg.fr/simbad/sim-basic?Ident=HIP034444&submit=SIMBAD+search}{34444} & 1.8 & 1.7 & \href{http://simbad.u-strasbg.fr/simbad/sim-basic?Ident=HIP060718&submit=SIMBAD+search}{60718} & 0.8 & 0.8 & \href{http://simbad.u-strasbg.fr/simbad/sim-basic?Ident=HIP078265&submit=SIMBAD+search}{78265} & 2.9 & 2.9 \\
\href{http://simbad.u-strasbg.fr/simbad/sim-basic?Ident=HIP015474&submit=SIMBAD+search}{15474} & 3.7 & 2.6 & \href{http://simbad.u-strasbg.fr/simbad/sim-basic?Ident=HIP034922&submit=SIMBAD+search}{34922} & 4.4 & 2.5 & \href{http://simbad.u-strasbg.fr/simbad/sim-basic?Ident=HIP060965&submit=SIMBAD+search}{60965} & 2.9 & 2.9 & \href{http://simbad.u-strasbg.fr/simbad/sim-basic?Ident=HIP078401&submit=SIMBAD+search}{78401} & 2.3 & 2.3 \\
\href{http://simbad.u-strasbg.fr/simbad/sim-basic?Ident=HIP015863&submit=SIMBAD+search}{15863} & 1.8 & 1.6 & \href{http://simbad.u-strasbg.fr/simbad/sim-basic?Ident=HIP035264&submit=SIMBAD+search}{35264} & 2.7 & 2.1 & \href{http://simbad.u-strasbg.fr/simbad/sim-basic?Ident=HIP061084&submit=SIMBAD+search}{61084} & 1.6 & 0.6 & \href{http://simbad.u-strasbg.fr/simbad/sim-basic?Ident=HIP078820&submit=SIMBAD+search}{78820} & 2.6 & 2.5 \\
\href{http://simbad.u-strasbg.fr/simbad/sim-basic?Ident=HIP017358&submit=SIMBAD+search}{17358} & 3.0 & 3.0 & \href{http://simbad.u-strasbg.fr/simbad/sim-basic?Ident=HIP035904&submit=SIMBAD+search}{35904} & 2.5 & 2.4 & \href{http://simbad.u-strasbg.fr/simbad/sim-basic?Ident=HIP061359&submit=SIMBAD+search}{61359} & 2.6 & 2.4 & \href{http://simbad.u-strasbg.fr/simbad/sim-basic?Ident=HIP079593&submit=SIMBAD+search}{79593} & 2.7 & 2.1 \\
\href{http://simbad.u-strasbg.fr/simbad/sim-basic?Ident=HIP017678&submit=SIMBAD+search}{17678} & 3.3 & 2.5 & \href{http://simbad.u-strasbg.fr/simbad/sim-basic?Ident=HIP036188&submit=SIMBAD+search}{36188} & 2.9 & 2.9 & \href{http://simbad.u-strasbg.fr/simbad/sim-basic?Ident=HIP061585&submit=SIMBAD+search}{61585} & 2.7 & 2.7 & \href{http://simbad.u-strasbg.fr/simbad/sim-basic?Ident=HIP079882&submit=SIMBAD+search}{79882} & 3.2 & 3.0 \\
\href{http://simbad.u-strasbg.fr/simbad/sim-basic?Ident=HIP017702&submit=SIMBAD+search}{17702} & 2.9 & 2.8 & \href{http://simbad.u-strasbg.fr/simbad/sim-basic?Ident=HIP036377&submit=SIMBAD+search}{36377} & 3.2 & 2.7 & \href{http://simbad.u-strasbg.fr/simbad/sim-basic?Ident=HIP061932&submit=SIMBAD+search}{61932} & 2.2 & 2.2 & \href{http://simbad.u-strasbg.fr/simbad/sim-basic?Ident=HIP080112&submit=SIMBAD+search}{80112} & 2.9 & 2.8 \\
\href{http://simbad.u-strasbg.fr/simbad/sim-basic?Ident=HIP018246&submit=SIMBAD+search}{18246} & 2.8 & 2.8 & \href{http://simbad.u-strasbg.fr/simbad/sim-basic?Ident=HIP036850&submit=SIMBAD+search}{36850} & 1.6 & 1.5 & \href{http://simbad.u-strasbg.fr/simbad/sim-basic?Ident=HIP061941&submit=SIMBAD+search}{61941} & 2.7 & 2.6 & \href{http://simbad.u-strasbg.fr/simbad/sim-basic?Ident=HIP080331&submit=SIMBAD+search}{80331} & 2.7 & 2.5 \\
\href{http://simbad.u-strasbg.fr/simbad/sim-basic?Ident=HIP018532&submit=SIMBAD+search}{18532} & 2.9 & 2.9 & \href{http://simbad.u-strasbg.fr/simbad/sim-basic?Ident=HIP037279&submit=SIMBAD+search}{37279} & 0.4 & 0.3 & \href{http://simbad.u-strasbg.fr/simbad/sim-basic?Ident=HIP062434&submit=SIMBAD+search}{62434} & 1.2 & 1.2 & \href{http://simbad.u-strasbg.fr/simbad/sim-basic?Ident=HIP080704&submit=SIMBAD+search}{80704} & 4.8 & 2.8 \\
\href{http://simbad.u-strasbg.fr/simbad/sim-basic?Ident=HIP018543&submit=SIMBAD+search}{18543} & 3.0 & 2.3 & \href{http://simbad.u-strasbg.fr/simbad/sim-basic?Ident=HIP037819&submit=SIMBAD+search}{37819} & 3.6 & 2.9 & \href{http://simbad.u-strasbg.fr/simbad/sim-basic?Ident=HIP062956&submit=SIMBAD+search}{62956} & 1.8 & 1.7 & \href{http://simbad.u-strasbg.fr/simbad/sim-basic?Ident=HIP080763&submit=SIMBAD+search}{80763} & 1.1 & -0.4 \\
\href{http://simbad.u-strasbg.fr/simbad/sim-basic?Ident=HIP021421&submit=SIMBAD+search}{21421} & 0.9 & 0.3 & \href{http://simbad.u-strasbg.fr/simbad/sim-basic?Ident=HIP037826&submit=SIMBAD+search}{37826} & 1.2 & 0.9 & \href{http://simbad.u-strasbg.fr/simbad/sim-basic?Ident=HIP063090&submit=SIMBAD+search}{63090} & 3.4 & 2.4 & \href{http://simbad.u-strasbg.fr/simbad/sim-basic?Ident=HIP080816&submit=SIMBAD+search}{80816} & 2.8 & 2.5 \\
\href{http://simbad.u-strasbg.fr/simbad/sim-basic?Ident=HIP021479&submit=SIMBAD+search}{21479} & 5.6 & 2.5 & \href{http://simbad.u-strasbg.fr/simbad/sim-basic?Ident=HIP039429&submit=SIMBAD+search}{39429} & 2.2 & 2.2 & \href{http://simbad.u-strasbg.fr/simbad/sim-basic?Ident=HIP063125&submit=SIMBAD+search}{63125} & 2.9 & 2.9 & \href{http://simbad.u-strasbg.fr/simbad/sim-basic?Ident=HIP081266&submit=SIMBAD+search}{81266} & 2.8 & 2.8 \\
\href{http://simbad.u-strasbg.fr/simbad/sim-basic?Ident=HIP023015&submit=SIMBAD+search}{23015} & 2.7 & 2.2 & \href{http://simbad.u-strasbg.fr/simbad/sim-basic?Ident=HIP039757&submit=SIMBAD+search}{39757} & 2.8 & 2.7 & \href{http://simbad.u-strasbg.fr/simbad/sim-basic?Ident=HIP063608&submit=SIMBAD+search}{63608} & 2.9 & 2.6 & \href{http://simbad.u-strasbg.fr/simbad/sim-basic?Ident=HIP081377&submit=SIMBAD+search}{81377} & 2.5 & 2.5 \\
\href{http://simbad.u-strasbg.fr/simbad/sim-basic?Ident=HIP023416&submit=SIMBAD+search}{23416} & 3.0 & 2.9 & \href{http://simbad.u-strasbg.fr/simbad/sim-basic?Ident=HIP039953&submit=SIMBAD+search}{39953} & 1.8 & 1.7 & \href{http://simbad.u-strasbg.fr/simbad/sim-basic?Ident=HIP064962&submit=SIMBAD+search}{64962} & 3.0 & 2.8 & \href{http://simbad.u-strasbg.fr/simbad/sim-basic?Ident=HIP081693&submit=SIMBAD+search}{81693} & 2.8 & 2.6 \\
\href{http://simbad.u-strasbg.fr/simbad/sim-basic?Ident=HIP023685&submit=SIMBAD+search}{23685} & 3.2 & 2.7 & \href{http://simbad.u-strasbg.fr/simbad/sim-basic?Ident=HIP041037&submit=SIMBAD+search}{41037} & 1.9 & 1.5 & \href{http://simbad.u-strasbg.fr/simbad/sim-basic?Ident=HIP065109&submit=SIMBAD+search}{65109} & 2.8 & 2.7 & \href{http://simbad.u-strasbg.fr/simbad/sim-basic?Ident=HIP082273&submit=SIMBAD+search}{82273} & 1.9 & 1.4 \\
\href{http://simbad.u-strasbg.fr/simbad/sim-basic?Ident=HIP023875&submit=SIMBAD+search}{23875} & 2.8 & 2.7 & \href{http://simbad.u-strasbg.fr/simbad/sim-basic?Ident=HIP042913&submit=SIMBAD+search}{42913} & 1.9 & 1.9 & \href{http://simbad.u-strasbg.fr/simbad/sim-basic?Ident=HIP065378&submit=SIMBAD+search}{65378} & 2.2 & 2.2 & \href{http://simbad.u-strasbg.fr/simbad/sim-basic?Ident=HIP082396&submit=SIMBAD+search}{82396} & 2.3 & 2.0 \\
\href{http://simbad.u-strasbg.fr/simbad/sim-basic?Ident=HIP024436&submit=SIMBAD+search}{24436} & 0.2 & 0.2 & \href{http://simbad.u-strasbg.fr/simbad/sim-basic?Ident=HIP043813&submit=SIMBAD+search}{43813} & 3.1 & 2.9 & \href{http://simbad.u-strasbg.fr/simbad/sim-basic?Ident=HIP065474&submit=SIMBAD+search}{65474} & 1.0 & 1.0 & \href{http://simbad.u-strasbg.fr/simbad/sim-basic?Ident=HIP082514&submit=SIMBAD+search}{82514} & 3.0 & 3.0 \\
\href{http://simbad.u-strasbg.fr/simbad/sim-basic?Ident=HIP024608&submit=SIMBAD+search}{24608} & 0.1 & -0.1 & \href{http://simbad.u-strasbg.fr/simbad/sim-basic?Ident=HIP044816&submit=SIMBAD+search}{44816} & 2.2 & 1.6 & \href{http://simbad.u-strasbg.fr/simbad/sim-basic?Ident=HIP066657&submit=SIMBAD+search}{66657} & 2.3 & 2.3 & \href{http://simbad.u-strasbg.fr/simbad/sim-basic?Ident=HIP083000&submit=SIMBAD+search}{83000} & 3.2 & 2.9 \\
\href{http://simbad.u-strasbg.fr/simbad/sim-basic?Ident=HIP025336&submit=SIMBAD+search}{25336} & 1.6 & 1.6 & \href{http://simbad.u-strasbg.fr/simbad/sim-basic?Ident=HIP045238&submit=SIMBAD+search}{45238} & 1.7 & 1.6 & \href{http://simbad.u-strasbg.fr/simbad/sim-basic?Ident=HIP067301&submit=SIMBAD+search}{67301} & 1.9 & 1.8 & -- & -- & -- \\
\href{http://simbad.u-strasbg.fr/simbad/sim-basic?Ident=HIP025428&submit=SIMBAD+search}{25428} & 1.6 & 1.6 & \href{http://simbad.u-strasbg.fr/simbad/sim-basic?Ident=HIP045556&submit=SIMBAD+search}{45556} & 2.2 & 2.1 & \href{http://simbad.u-strasbg.fr/simbad/sim-basic?Ident=HIP067457&submit=SIMBAD+search}{67457} & 4.2 & 2.7 & -- & -- & -- \\

\hline
\end{tabular}
\end{center}
\end{table}

\begin{table}[h!]
\caption{Continuation of Table \ref{tab:vbsstars}.}\vspace{1mm}
\label{tab:vbsstars2}
\begin{center}
\begin{tabular}{rrr|rrr|rrr|rrr}
\hline
\hline
HIP & $V$ & $G$ & HIP & $V$ & $G$ & HIP & $V$ & $G$ & HIP & $V$ & $G$ \\
   & (mag) & (mag) &    & (mag) & (mag) &    & (mag) & (mag) &    & (mag) & (mag) \\ 
\hline
\href{http://simbad.u-strasbg.fr/simbad/sim-basic?Ident=HIP083081&submit=SIMBAD+search}{83081} & 3.1 & 2.6 & \href{http://simbad.u-strasbg.fr/simbad/sim-basic?Ident=HIP087833&submit=SIMBAD+search}{87833} & 2.2 & 1.7 & \href{http://simbad.u-strasbg.fr/simbad/sim-basic?Ident=HIP095947&submit=SIMBAD+search}{95947} & 3.0 & 2.8 & \href{http://simbad.u-strasbg.fr/simbad/sim-basic?Ident=HIP107315&submit=SIMBAD+search}{107315} & 2.4 & 1.9 \\
\href{http://simbad.u-strasbg.fr/simbad/sim-basic?Ident=HIP084012&submit=SIMBAD+search}{84012} & 2.4 & 2.4 & \href{http://simbad.u-strasbg.fr/simbad/sim-basic?Ident=HIP088635&submit=SIMBAD+search}{88635} & 3.0 & 2.7 & \href{http://simbad.u-strasbg.fr/simbad/sim-basic?Ident=HIP097165&submit=SIMBAD+search}{97165} & 2.9 & 2.8 & \href{http://simbad.u-strasbg.fr/simbad/sim-basic?Ident=HIP107556&submit=SIMBAD+search}{107556} & 2.9 & 2.8 \\
\href{http://simbad.u-strasbg.fr/simbad/sim-basic?Ident=HIP084345&submit=SIMBAD+search}{84345} & 2.8 & 2.5 & \href{http://simbad.u-strasbg.fr/simbad/sim-basic?Ident=HIP089642&submit=SIMBAD+search}{89642} & 3.1 & 2.2 & \href{http://simbad.u-strasbg.fr/simbad/sim-basic?Ident=HIP097278&submit=SIMBAD+search}{97278} & 2.7 & 2.3 & \href{http://simbad.u-strasbg.fr/simbad/sim-basic?Ident=HIP108085&submit=SIMBAD+search}{108085} & 3.0 & 3.0 \\
\href{http://simbad.u-strasbg.fr/simbad/sim-basic?Ident=HIP084380&submit=SIMBAD+search}{84380} & 3.2 & 2.8 & \href{http://simbad.u-strasbg.fr/simbad/sim-basic?Ident=HIP089931&submit=SIMBAD+search}{89931} & 2.7 & 2.3 & \href{http://simbad.u-strasbg.fr/simbad/sim-basic?Ident=HIP097649&submit=SIMBAD+search}{97649} & 0.8 & 0.7 & \href{http://simbad.u-strasbg.fr/simbad/sim-basic?Ident=HIP109074&submit=SIMBAD+search}{109074} & 3.0 & 2.7 \\
\href{http://simbad.u-strasbg.fr/simbad/sim-basic?Ident=HIP085258&submit=SIMBAD+search}{85258} & 2.8 & 2.3 & \href{http://simbad.u-strasbg.fr/simbad/sim-basic?Ident=HIP089962&submit=SIMBAD+search}{89962} & 3.2 & 3.0 & \href{http://simbad.u-strasbg.fr/simbad/sim-basic?Ident=HIP098337&submit=SIMBAD+search}{98337} & 3.5 & 2.9 & \href{http://simbad.u-strasbg.fr/simbad/sim-basic?Ident=HIP109268&submit=SIMBAD+search}{109268} & 1.7 & 1.7 \\
\href{http://simbad.u-strasbg.fr/simbad/sim-basic?Ident=HIP085670&submit=SIMBAD+search}{85670} & 2.8 & 2.5 & \href{http://simbad.u-strasbg.fr/simbad/sim-basic?Ident=HIP090185&submit=SIMBAD+search}{90185} & 1.8 & 1.8 & \href{http://simbad.u-strasbg.fr/simbad/sim-basic?Ident=HIP100345&submit=SIMBAD+search}{100345} & 3.0 & 2.8 & \href{http://simbad.u-strasbg.fr/simbad/sim-basic?Ident=HIP109492&submit=SIMBAD+search}{109492} & 3.4 & 2.8 \\
\href{http://simbad.u-strasbg.fr/simbad/sim-basic?Ident=HIP085696&submit=SIMBAD+search}{85696} & 2.7 & 2.7 & \href{http://simbad.u-strasbg.fr/simbad/sim-basic?Ident=HIP090496&submit=SIMBAD+search}{90496} & 2.8 & 2.5 & \href{http://simbad.u-strasbg.fr/simbad/sim-basic?Ident=HIP100453&submit=SIMBAD+search}{100453} & 2.2 & 2.1 & \href{http://simbad.u-strasbg.fr/simbad/sim-basic?Ident=HIP110130&submit=SIMBAD+search}{110130} & 2.9 & 2.4 \\
\href{http://simbad.u-strasbg.fr/simbad/sim-basic?Ident=HIP085792&submit=SIMBAD+search}{85792} & 2.8 & 2.8 & \href{http://simbad.u-strasbg.fr/simbad/sim-basic?Ident=HIP091262&submit=SIMBAD+search}{91262} & 0.0 & 0.0 & \href{http://simbad.u-strasbg.fr/simbad/sim-basic?Ident=HIP100751&submit=SIMBAD+search}{100751} & 1.9 & 1.9 & \href{http://simbad.u-strasbg.fr/simbad/sim-basic?Ident=HIP111043&submit=SIMBAD+search}{111043} & 4.1 & 3.0 \\
\href{http://simbad.u-strasbg.fr/simbad/sim-basic?Ident=HIP085927&submit=SIMBAD+search}{85927} & 1.6 & 1.6 & \href{http://simbad.u-strasbg.fr/simbad/sim-basic?Ident=HIP092855&submit=SIMBAD+search}{92855} & 2.0 & 2.0 & \href{http://simbad.u-strasbg.fr/simbad/sim-basic?Ident=HIP101772&submit=SIMBAD+search}{101772} & 3.1 & 2.8 & \href{http://simbad.u-strasbg.fr/simbad/sim-basic?Ident=HIP112122&submit=SIMBAD+search}{112122} & 2.1 & 0.9 \\
\href{http://simbad.u-strasbg.fr/simbad/sim-basic?Ident=HIP086032&submit=SIMBAD+search}{86032} & 2.1 & 2.0 & \href{http://simbad.u-strasbg.fr/simbad/sim-basic?Ident=HIP092862&submit=SIMBAD+search}{92862} & 4.1 & 2.4 & \href{http://simbad.u-strasbg.fr/simbad/sim-basic?Ident=HIP102098&submit=SIMBAD+search}{102098} & 1.2 & 1.2 & \href{http://simbad.u-strasbg.fr/simbad/sim-basic?Ident=HIP112158&submit=SIMBAD+search}{112158} & 2.9 & 2.7 \\
\href{http://simbad.u-strasbg.fr/simbad/sim-basic?Ident=HIP086228&submit=SIMBAD+search}{86228} & 1.9 & 1.8 & \href{http://simbad.u-strasbg.fr/simbad/sim-basic?Ident=HIP093506&submit=SIMBAD+search}{93506} & 2.6 & 2.6 & \href{http://simbad.u-strasbg.fr/simbad/sim-basic?Ident=HIP102488&submit=SIMBAD+search}{102488} & 2.5 & 2.2 & \href{http://simbad.u-strasbg.fr/simbad/sim-basic?Ident=HIP112961&submit=SIMBAD+search}{112961} & 3.7 & 2.9 \\
\href{http://simbad.u-strasbg.fr/simbad/sim-basic?Ident=HIP086670&submit=SIMBAD+search}{86670} & 2.4 & 2.4 & \href{http://simbad.u-strasbg.fr/simbad/sim-basic?Ident=HIP093747&submit=SIMBAD+search}{93747} & 3.0 & 3.0 & \href{http://simbad.u-strasbg.fr/simbad/sim-basic?Ident=HIP104732&submit=SIMBAD+search}{104732} & 3.2 & 3.0 & \href{http://simbad.u-strasbg.fr/simbad/sim-basic?Ident=HIP113368&submit=SIMBAD+search}{113368} & 1.2 & 1.1 \\
\href{http://simbad.u-strasbg.fr/simbad/sim-basic?Ident=HIP086742&submit=SIMBAD+search}{86742} & 2.8 & 2.4 & \href{http://simbad.u-strasbg.fr/simbad/sim-basic?Ident=HIP093864&submit=SIMBAD+search}{93864} & 3.3 & 3.0 & \href{http://simbad.u-strasbg.fr/simbad/sim-basic?Ident=HIP105199&submit=SIMBAD+search}{105199} & 2.5 & 2.4 & \href{http://simbad.u-strasbg.fr/simbad/sim-basic?Ident=HIP113881&submit=SIMBAD+search}{113881} & 2.4 & 1.4 \\
\href{http://simbad.u-strasbg.fr/simbad/sim-basic?Ident=HIP087073&submit=SIMBAD+search}{87073} & 3.0 & 2.8 & \href{http://simbad.u-strasbg.fr/simbad/sim-basic?Ident=HIP094141&submit=SIMBAD+search}{94141} & 2.9 & 2.8 & \href{http://simbad.u-strasbg.fr/simbad/sim-basic?Ident=HIP106278&submit=SIMBAD+search}{106278} & 2.9 & 2.7 & \href{http://simbad.u-strasbg.fr/simbad/sim-basic?Ident=HIP113963&submit=SIMBAD+search}{113963} & 2.5 & 2.5 \\
\href{http://simbad.u-strasbg.fr/simbad/sim-basic?Ident=HIP087261&submit=SIMBAD+search}{87261} & 3.2 & 2.9 & \href{http://simbad.u-strasbg.fr/simbad/sim-basic?Ident=HIP094376&submit=SIMBAD+search}{94376} & 3.1 & 2.8 & \href{http://simbad.u-strasbg.fr/simbad/sim-basic?Ident=HIP107259&submit=SIMBAD+search}{107259} & 4.2 & 2.2 & \href{http://simbad.u-strasbg.fr/simbad/sim-basic?Ident=HIP116727&submit=SIMBAD+search}{116727} & 3.2 & 2.9 \\

\hline
\end{tabular}
\end{center}
\end{table}

\end{appendix}

\end{document}